\documentclass[11pt]{article}

\PassOptionsToPackage{table}{xcolor}
\usepackage[final]{acl}

\usepackage{times}
\usepackage{latexsym}
\usepackage[T1]{fontenc}
\usepackage[utf8]{inputenc}
\usepackage{microtype}
\usepackage{inconsolata}
\usepackage{graphicx}
\usepackage{booktabs}
\usepackage{enumitem}
\usepackage{amssymb}
\usepackage{amsmath}
\usepackage{pifont}
\usepackage{subcaption}

\definecolor{darkgreen}{rgb}{0,0.5,0}

\newcommand{\xmark}{\textcolor{red}{\ding{55}}}
\newcommand{\cmark}{\textcolor{darkgreen}{\ding{51}}}

\title{Doc-REFRAG: Rethinking Multimodal Document Retrieval-Augmented Generation}

\author{
  \bfseries Ruofan Hu\thanks{Equal contribution.},
  Shengyang Xu\footnotemark[1],
  Minjie Hong,
  Xiaoda Yang,
\\
  \bfseries Sashuai Zhou,          
  Ke Lei,
  Tao Jin,
  Zhou Zhao\thanks{Corresponding author.}
\\
  Zhejiang University
\\
  \texttt{ruofanhu@zju.edu.cn}
}

\begin{document}

\maketitle

\begin{abstract}
Real-world knowledge resides in multimodal documents, necessitating retrieval-augmented generation (RAG) for accurate question answering. However, existing multimodal RAG models are primarily designed for single-image or closed-document settings and exhibit limited accuracy in realistic multi-image scenarios. Moreover, processing numerous retrieved images incurs substantial computational overhead from irrelevant visual tokens. To address these challenges, we introduce DocLongRAG, a large-scale dataset of 343K question--answer pairs, each associated with an average of 37.4 retrieved images to reflect authentic RAG workflows. Building on this dataset, we propose Doc-REFRAG, a question-guided framework that compresses visual tokens into coarse chunks and selectively expands question-relevant ones via a lightweight RL-based selector. Experiments on six benchmarks show that Doc-REFRAG outperforms eleven strong baselines, achieving state-of-the-art accuracy with significantly lower inference latency. Our resources are available at \url{https://github.com/Collab-Gen/Doc-REFRAG}.
\end{abstract}

\section{Introduction}
Multimodal large language models (MLLMs) have rapidly developed, achieving remarkable performance in various visual understanding tasks~\cite{lu2024deepseek, bai2025qwen2, chen2024internvl,yao2024minicpm,li2024llava}. Despite these advances, current MLLMs remain limited by static parametric knowledge. To address this limitation, RAG enhances MLLMs with retrieved evidence from external corpora, thereby expanding factual coverage and mitigating hallucination~\cite{arslan2024survey}. Crucially, a significant portion of real-world knowledge resides in visually rich documents, which often contain textual passages, charts, and figures. However, existing MLLMs trained on individual images struggle with multi-image reasoning~\cite{ma2024mmlongbench, dong2025mmdocir, tito2023hierarchical}. Moreover, substantial visual tokens from retrieved images impose significant computational overhead~\cite{ye2023mplug, liu2024textmonkey}.

\begin{figure}[t!]
    \centering
    \includegraphics[width=\columnwidth,trim=18mm 2mm 10mm 0mm, keepaspectratio, page=1, clip]{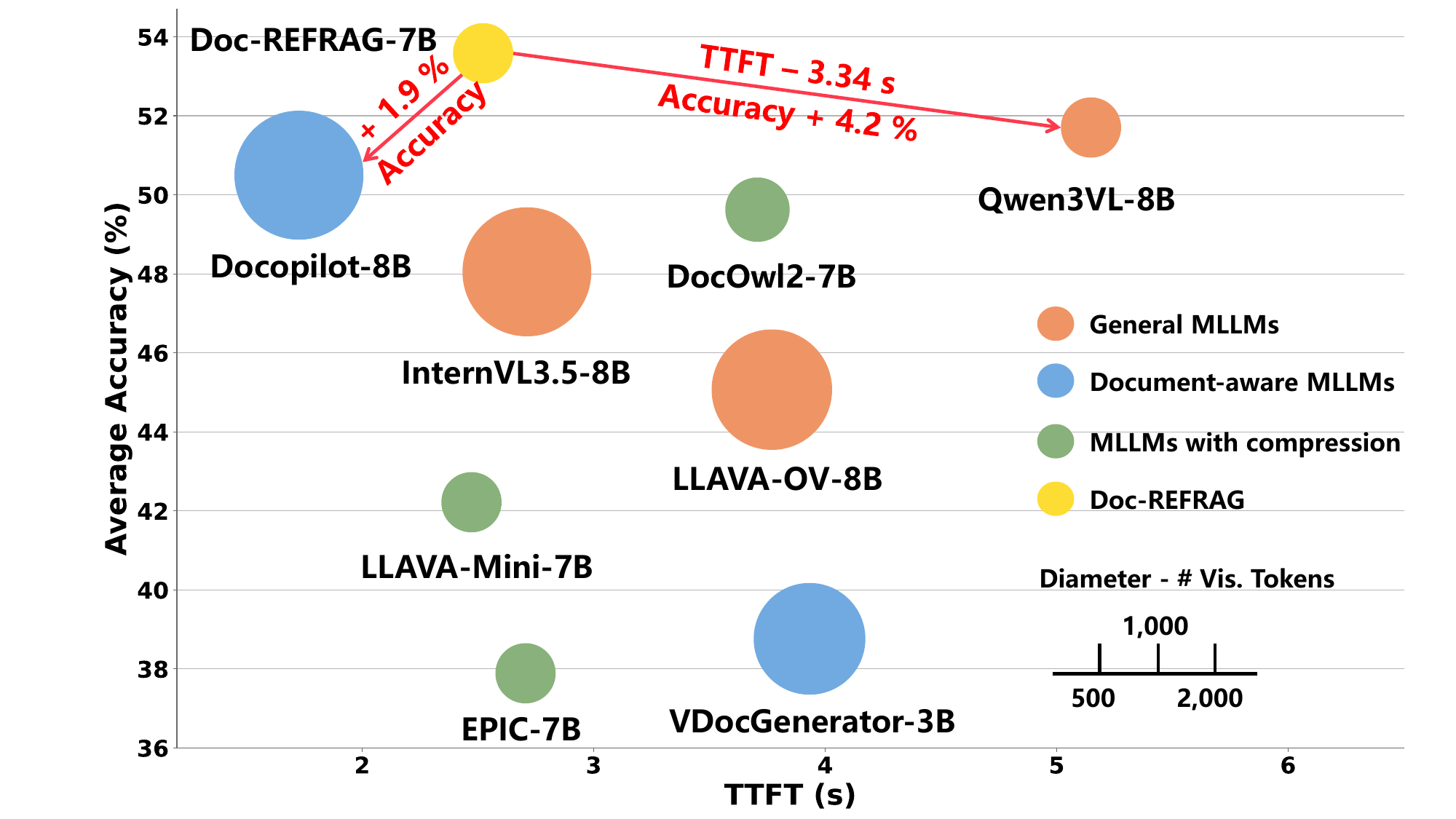}
    \caption{Doc-REFRAG achieves the best accuracy--efficiency trade-off among MLLMs.}
    \vspace{-14pt}
    \label{fig:intro_fig}
\end{figure}

To handle multi-image inputs, recent works construct long-document training corpora~\cite{van2023document, docowl15, tanaka2025vdocrag, duan2025docopilot} and apply curriculum learning to extend capacity from single to multi-image layouts~\cite{docowl2, duan2025docopilot}. However, these works generally assume a closed-domain setting where all images belong to a single, coherent document. RAG, however, retrieves images across diverse documents that are heterogeneous, redundant, and contradictory. As shown in~\cite{dong2025benchmarking, cuconasu2024power}, this distribution shift severely degrades MLLM generalization to the fragmented and noisy images encountered during RAG inference.

Orthogonal to the reasoning challenge, existing compression methods fall into two types. Training-free approaches typically rely on either token importance~\cite{chen2024image, he2024zipvl} or redundancy estimation~\cite{tan2025tokencarve, wang2025folder} to prune visual tokens. In contrast, training-aware methods introduce a linear projector to refine the limited coarse tokens with fine-grained visual details~\cite{li2025tokenpacker,docowl2}. Both categories prioritize visually complex patches, assuming such regions are more informative for downstream tasks. This assumption fails in RAG scenarios, where visually complex patches are often question-irrelevant. Consequently, processing multiple retrieved images still incurs numerous redundant visual tokens and substantial prefill latency~\cite{li2024survey}.

To address these problems, we introduce DocLongRAG, a new multimodal document dataset. Compared to existing work~\cite{docowl15,tanaka2025vdocrag,duan2025docopilot,luo2025towards}, this dataset has the following features. (1) \textbf{High-Quality Retrieval Contexts}: semantically similar images are explicitly interleaved with hard negative distractors to simulate real-world RAG inputs. (2) \textbf{Long-Range Reasoning}: each question is associated with an average of 37.4 images, enabling training on substantially longer sequences. (3) \textbf{Broad Format Coverage}: 343,474 question--answer pairs grounded in 3.3M document images spanning PDFs, slides, and posters, supporting robust generalization across document formats.

Building on DocLongRAG, we propose Doc-REFRAG, a framework for efficient multimodal RAG decoding. Unlike pruning-based methods, Doc-REFRAG adopts a ``compress-and-expand'' strategy: visual tokens are first compressed into non-overlapping chunks. A lightweight question-conditioned selector then re-expands only the most relevant chunks into full-resolution tokens, recovering fine-grained semantics precisely where needed without the context loss of static pruning.

Integrating mixed sequences of chunked and original tokens requires a specialized training strategy. Doc-REFRAG adopts a three-stage framework: single-image reconstruction for chunk semantic fidelity, multi-image continual pretraining for mixed-sequence processing, and RL-based selector training using accuracy as reward. As shown in Fig.~\ref{fig:intro_fig}, this cuts response time while improving accuracy over prior work.

The contributions are summarized as follows:

\begin{itemize}[leftmargin=*,itemsep=1pt,topsep=2pt]
    \item We introduce DocLongRAG, a large-scale multimodal dataset reflecting the noisy, long-context nature of real-world retrieval.

    \item We propose Doc-REFRAG, which rethinks visual token compression as dynamic, question-guided selective expansion, via a three-stage curriculum with an RL-based selector.

    \item Experiments on six benchmarks show that our method outperforms eleven existing approaches.
\end{itemize}

\section{Related Work}
\subsection{Document Understanding Models in RAG}
RAG enhances MLLMs by retrieving external knowledge. In multimodal document RAG, existing research focuses on retriever design~\cite{yu2024visrag,yan2025docpruner,zhang2025towards,hu2025vela,yang2025multimodal,hu2026docretriever} while employing general-purpose MLLMs as generators. Early VDU MLLMs~\cite{appalaraju2024docformerv2,wang2023docllm} and subsequent multi-image variants~\cite{feng2024docpedia,duan2025docopilot,liu2025knowledge} primarily target coherent, single-source documents. In contrast, real-world RAG involves retrieving fragmented and contradictory images from diverse sources, creating a distribution shift that current models struggle to handle. While some works reason over massive document piles \cite{chen2025document}, they overlook the efficiency required for interactive generation.

\begin{figure*}[t!]
    \centering
    \begin{minipage}[b]{0.48\textwidth}
        \centering
        \includegraphics[width=\linewidth, trim=36mm 48mm 25mm 40mm, clip]{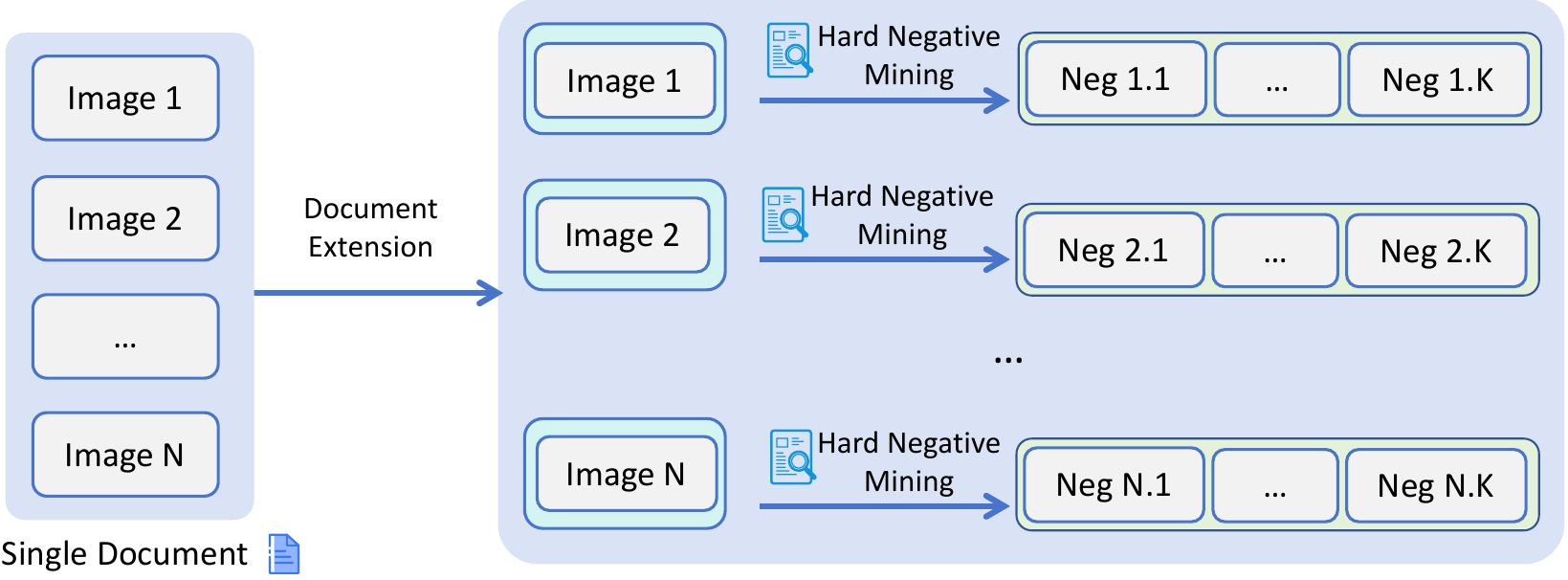}
        \vspace{-10pt}
        \caption{Process of creating DocLongRAG.}
        \label{fig:multihop}
    \end{minipage}
    \hfill
    \begin{minipage}[b]{0.5\textwidth}
        \centering
        \resizebox{\linewidth}{!}{
        \begin{tabular}{lccc}
            \toprule
             & Doc-750k & OpenDocVQA & DocLongRAG \\ \midrule
            Context-Dependent & \xmark & \cmark & \cmark  \\
            Online-Independent & \xmark  & \cmark& \cmark \\
            Answerable  & \cmark  & \xmark & \cmark \\
            Ethically Reviewed & \xmark  & \xmark & \cmark \\
            Retrieval-Noisy & \xmark  & \xmark & \cmark \\
            \#QA Pairs & 758,000 & 43,474 & 343,474 \\
            \#Images (Pages) & 3,100,000 & 206,267 & 3,306,267 \\
            \#Avg. Images per Question & 4.1 & 2.1 & 37.4 \\
            \bottomrule
        \end{tabular}
        }
        \vspace{-4pt}
        \captionof{table}{Comparison of related datasets.}
        \label{tab:comparison}
    \end{minipage}
    \vspace{-12pt}
\end{figure*}

\subsection{Visual Token Compression}
Multi-image inputs cause a token explosion that degrades inference efficiency, motivating visual token compression. These methods fall into training-free and training-aware approaches. Training-free methods typically adopt importance-based strategies~\cite{chen2024image,zhang2024sparsevlm}, pruning tokens by attention scores, and redundancy-based approaches~\cite{jiang2025kind,wen2025token}, merging similar tokens by feature similarity. Training-aware methods adopt lightweight component replacements~\cite{dai2023instructblip,dong2024internlm,li2024llama}. Recent examples include MQT-LLaVA~\cite{hu2024matryoshka}, which uses a dynamic Q-Former, and TokenPacker~\cite{li2025tokenpacker} and DocOwl2~\cite{docowl2}, which use coarse-to-fine condensation. However, these methods prioritize visually salient regions~\cite{zhang2025llava}, misaligning with question relevance in RAG. A structured comparison across six design dimensions is in App.~\ref{appendix:differentiation}.

\section{DocLongRAG}
\label{sec:dataset}

Conventional VDU datasets focus on individual or coherent images, offering limited context. Conversely, RAG requires reasoning across long, noisy, and conflicting retrieved inputs. DocLongRAG simulates the scale and noise of practical retrieval, engineered for training stability.

\subsection{Dataset Collection and Filtering}
To support RAG training, we curate a diverse VDU corpus integrating Doc-750K~\cite{duan2025docopilot} and OpenDocVQA~\cite{tanaka2025vdocrag}, which aggregate English data from 15 open-source datasets (see App.~\ref{appendix:source_datasets}) and scientific documents. To address instances conflicting with RAG requirements or ethical guidelines, we filter four sample categories:

\begin{itemize}[left=0pt]
    \item \textbf{Context-independent questions}: Factoid questions answerable from parametric knowledge alone, without any retrieved document image (e.g., ``Who won the Turing Award in 2018?'').
    
    \item \textbf{Online search questions}: Questions requiring external web resources beyond the document context (e.g., ``What is the author's Google Scholar citation count?''), violating the assumption that all information resides in the documents.
    
    \item \textbf{Unanswerable questions}: Questions lacking factual grounding in the documents, excluded as they do not support document-based reasoning.
    
    \item \textbf{Ethically problematic content}: Questions with personally identifiable information, offensive language, harmful instructions, or sensitive imagery, violating ethical standards.
\end{itemize}

This filtering reduces the pool by 18.2\%, yielding a curated corpus $C$ enforcing retrieval dependence. The corpus retains diverse formats (PDFs, web screenshots, slides) and answer types (extractive, abstractive), providing a challenging training foundation. This filtering applies only to the training corpus, while the evaluation benchmarks retain unanswerable and context-independent questions.

\subsection{Negative Document Extension}
We construct DocLongRAG to train VDU models on multiple document images under realistic retrieval noise. Inspired by~\cite{gao2025nextlong}, we interleave relevant images with hard-negative distractors from $C$ to form extended input sequences.

Specifically, each instance in $C$ comprises a question $q$, relevant document images $D = \{d_1, \ldots, d_N\}$, and an answer $a$. As shown in Fig.~\ref{fig:multihop}, for each image $d_j \in D$, we retrieve the top-$k$ most similar images from $C \setminus \mathcal{D}$ using the document-aware retriever~\cite{gunther2025jina}. However, due to the semantic richness of document content, some images $\in C \setminus \mathcal{D}$ may still contain answer-supporting evidence. To avoid mislabeling such ambiguous candidates as negatives, we manually evaluate 500 randomly sampled queries and find that on average the top 2.47 retrieved images remain relevant (App.~\ref{appendix:hard_negative}). We therefore exclude the top-3 results and define the next $m$ images (ranked 4 onward) as hard negatives for $d_j$, where $m$ is uniformly sampled from $[20, 30]$. We construct the final sequence by interleaving each relevant image $d_j$ with its hard negatives as $l = \big[\, \{d_j, n_j^1, \ldots, n_j^m\}_{j=1}^N \,\big]$. This yields a long, redundant, and sometimes conflicting multimodal context, closely mirroring real-world RAG. Applying this to all instances in $C$ yields the DocLongRAG corpus. We emphasize that such constructed noise is used only for training, while evaluation relies on the noise produced by a real retriever over real benchmark documents (Sec.~\ref{sec:experiments}).

Through ablation studies on the placement position of relevant images (i.e., the meta-chunk) within the hard-negative sequence (App.~\ref{appendix:metachunkposition}), we show that placing relevant document images before hard negatives consistently improves model performance during training. This strategic ordering helps establish longer dependencies, enhancing the effectiveness of DocLongRAG.

\subsection{Comparison with Related Datasets}
Tab.~\ref{tab:comparison} presents a statistical comparison of DocLongRAG with existing VDU datasets. DocLongRAG differs from prior work along three key dimensions, establishing a more challenging and realistic setting for training. First, it aggregates instances from multiple source datasets, covering diverse document image formats and question types. Second, while Doc-750K contains context-independent questions and OpenDocVQA comprises unanswerable ones, DocLongRAG requires all answers to be grounded in the provided document images, enforcing a retrieval-dependent setting. Third, whereas OpenDocVQA grounds questions in individual images and Doc-750K in coherent document images, DocLongRAG associates each question with an average of 37.4 images, intentionally curated to include redundancy, noise, and conflicting evidence.

\section{Doc-REFRAG}
\subsection{Model Architecture}
\begin{figure*}
    \centering
    \includegraphics[width=\textwidth, height=6cm, keepaspectratio,trim=85mm 18mm 80mm 30mm, page=1, clip]{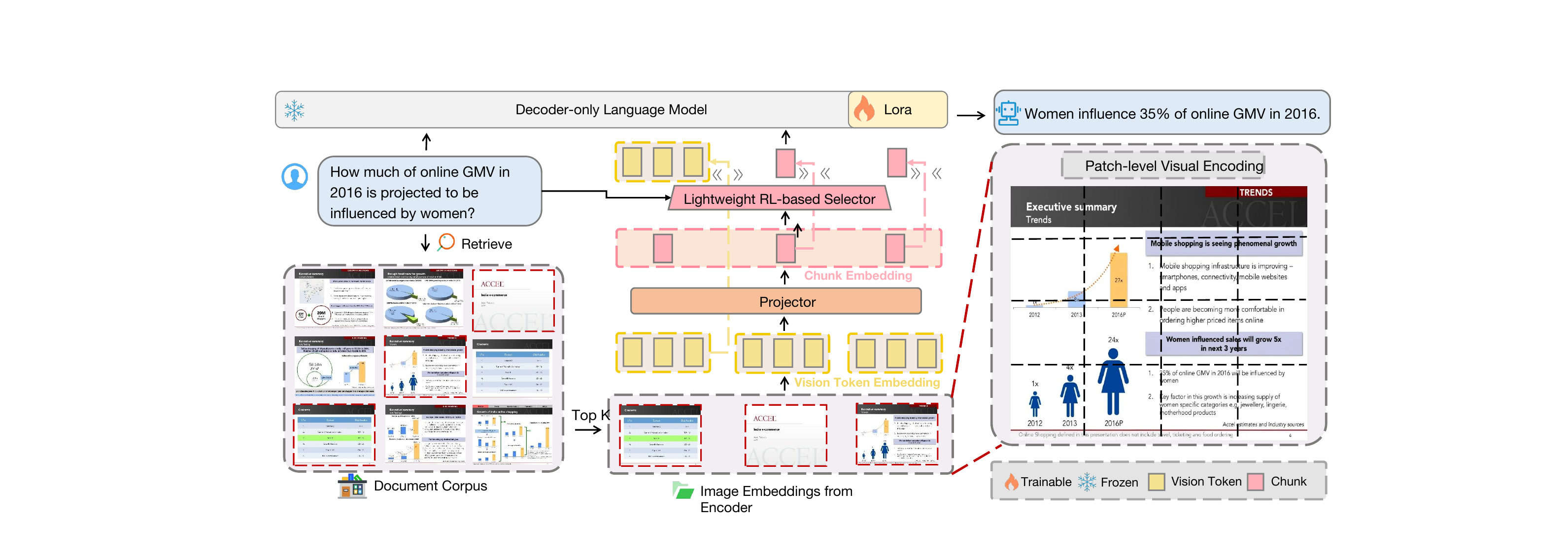}
    \caption{Architecture of Doc-REFRAG. Retrieved document images are encoded patch-level into chunk embeddings. A query-aware RL-based selector identifies informative chunks, which are expanded to fine-grained tokens before being fed with the query into a decoder-only language model.}
    \label{fig:main}
    \vspace{-12pt}
\end{figure*}

The overall architecture of Doc-REFRAG is illustrated in Fig.~\ref{fig:main}, consisting of four key components: (1) a document-tailored vision encoder built upon DocOwl2, (2) a two-layer MLP projector for chunk embedding, (3) a lightweight RL-based selector for dynamic token restoration, and (4) a decoder-only language model. A consolidated overview of all components, training stages, and inference-time operations is provided in App.~\ref{appendix:system_overview}.

Given a question consisting of $q$ tokens, denoted as $\mathbf{t} = [t_1, \dots, t_q]$, each retrieved document image is encoded by the vision encoder, which maps the input image to a compact sequence of $m = 324$ visual tokens. Concatenating these visual tokens from all $n$ retrieved images yields the initial visual token sequence in the input  $\mathbf{v} = [v_1, \dots, v_{n\times m}]$.

In RAG, only a small subset of regions is semantically relevant to the question, rendering most visual tokens redundant. Existing compression methods, however, are typically question-agnostic: they prioritize visually dense regions (e.g., text-heavy blocks) and discard sparser ones (e.g., blank areas), yet visual density correlates poorly with semantic relevance.

To address this mismatch, we propose a question-guided compression strategy that prioritizes semantic relevance over visual appearance. Specifically, Doc-REFRAG first partitions the visual token sequence of each retrieved image into $L = m / k$ non-overlapping chunks, which serve as units for coarse-level semantic modeling. For the $i$-th image, we define its chunk group $\mathcal{G}_{(i)} = \{ C_i^1, \dots, C_i^L \}$, where each chunk $C_i^j$ consists of $k$ tokens from $v_i$. This chunking preserves the 2D arrangement of the page. We chunk layout-aware tokens rather than raw patches: each token from the encoder summarizes a fixed region of the page and has aggregated global context through self-attention. Since consecutive tokens in raster order are spatially adjacent, chunking them keeps each group localized on the page rather than scattered across arbitrary positions. Expansion then restores the original tokens of exactly the same region, reinstating fine-grained spatial detail where the selector fires. In this way, two-dimensional reasoning is inherited from the encoder, while our compression reuses rather than rebuilds this spatial structure.

A two-layer MLP projector $\phi$ then projects each chunk into a single $d$-dimensional embedding $\mathbf{c}_i^j = \phi(C_i^j) \in \mathbb{R}^{d}$, where $d$ denotes the hidden dimension of the decoder's input embedding space, preserving semantic content while reducing length. The chunk embeddings from all retrieved images are concatenated with question tokens to form the input sequence:
\[
\mathbf{z} = [t_1, \dots, t_q, \mathbf{c}_1^1, \dots, \mathbf{c}_n^L] \in \mathbb{R}^{(q + n \times L) \times d},
\]
which is fed into a decoder-only foundation model to generate the answer $y \sim p_\theta(y \mid \mathbf{z})$.

To retain fine-grained details that coarse embeddings may obscure, a lightweight RL-based selector dynamically identifies critical chunks and expands them back to their original $k$ constituent tokens during decoding. Let $\delta_i^j \in \{0, 1\}$ denote whether the $j$-th chunk of the $i$-th image should be expanded. The final input sequence becomes $\mathbf{z}' = [t_1, \dots, t_q, \tilde{\mathbf{c}}_1^1, \dots, \tilde{\mathbf{c}}_n^L]$, where $\tilde{\mathbf{c}}_i^j = C_i^j \in \mathbb{R}^{k \times d}$ if $\delta_i^j = 1$ and $\tilde{\mathbf{c}}_i^j = \mathbf{c}_i^j \in \mathbb{R}^{d}$ otherwise. This mechanism balances computational efficiency and semantic fidelity based on the question.

\subsection{Model Training}
To align the encoder and decoder with $\mathbf{z}'$, we adopt a three-stage protocol: Single-Image Reconstruction, Multi-image Continual Pretraining (CPT), and RL-Based Selector Training.

\paragraph{Single-Image Reconstruction.}
We freeze the decoder and train the encoder and MLP projector on DocStruct4M~\cite{docowl15} with two objectives: (1) aligning chunk embeddings with the decoder's input embedding space, and (2) preserving original visual semantics. We formulate two complementary tasks: text parsing, where the model recognizes all textual content in a document image, and text localization, where it extracts text from a spatially specified region. The encoder is then frozen for subsequent stages.

\paragraph{Multi-image Continual Pretraining.}
In the CPT stage, we employ curriculum learning to train the decoder on increasingly complex datasets, with two objectives: (1) scaling from single- to multi-image settings, and (2) enabling effective processing of hybrid sequences that interleave chunk and visual token embeddings.

Chunk expansion is supervised by pseudo-labels $\delta_i^j$ derived from ColQwen~\cite{faysse2024colpali} question-to-patch similarity scores. Since our encoder and ColQwen share the same normalized image coordinate convention, a deterministic geometric mapping from each coarse chunk to the fine patches it subsumes is established (App.~\ref{appendix:geometric_mapping}; robustness in App.~\ref{appendix:pseudo_label_robustness}). At inference time, $\delta_i^j$ is predicted by the RL-based selector.

Building upon this pseudo-labeling strategy, we begin with OpenDocVQA~\cite{tanaka2025vdocrag}, a single-image VDU dataset: for each question, we expand the top 20\% of chunks ranked by ColQwen similarity scores, enabling the decoder to fuse fine-grained visual details with coarse chunk summaries within a single image.

\begin{table*}
    \footnotesize
    \centering
    \resizebox{\linewidth}{!}{
    \begin{tabular}{lcc|cccccc|ccc}
    \toprule
     \textbf{Model} & \textbf{Size} & \textbf{Token$^V$} &
     \textbf{ChartQA} &
     \textbf{SlideVQA} &
     \textbf{InfoVQA} &
     \textbf{DUDE} &
     \textbf{Vidore} &
     \textbf{MMDocIR} &
     \textbf{\#Acc.} & \textbf{\#TTFT} & \textbf{\#TTIT} \\
    \midrule
    \multicolumn{12}{l}{\textbf{General MLLMs}} \\
    InternVL3.5 & 8B & $\sim$2,304 &
    56.8 &	53.1 &	48.1 &	45.8 &	42.1 &	51.5 &	49.5  & 2.78 & 0.92 \\
    LLaVA-OneVision & 8B & $\sim$1,920 &
    52.1 &	50.9 &	46.3 &	47.2 &	42.7 &	49.0 &	48.0  & 4.03 & \textbf{0.05} \\
    MiniCPM-V-2.6 & 8B & $\sim$576 &
    54.8 & 52.3 & 48.5 & 46.4 & 41.1 & 48.8 & 48.6 & 5.80 & 0.12 \\
    Qwen3-VL & 8B & $\sim$512 &
    58.2 & 59.2 & 55.7 & 52.5 & 43.5 & 52.4 & 51.3 & 5.91 & \textbf{0.05} \\
    MQT-LLaVA$^\dagger$ & 7B & 256 &
    35.7 & 39.3 & 33.4 & 40.5 & 37.4 & 34.2 & 36.7 & 3.43 & \textbf{0.05} \\
    EPIC$^\dagger$ & 7B & 256 &
    39.7 & 41.1 & 42.7 & 32.4 & 33.3 & 38.2 & 37.9 & 3.42 & \underline{0.06} \\
    LLaVA-Mini$^\dagger$ & 7B & 144 &
    35.4 & 38.5 & 44.6 & 37.5 & 39.1 & 40.2 & 39.2 & 2.67 & \underline{0.06} \\
    \midrule
    \multicolumn{12}{l}{\textbf{Document-aware MLLMs}} \\
    Docopilot & 8B & $\sim$2304 &
    \underline{58.7} & \underline{56.6} & 49.4 & 48.6 & 45.7 & 46.5 & 50.9 & \textbf{1.81} & 2.25 \\
    VDocGenerator & 3B & $\sim$1920 &
    53.3 & 42.8 & 43.2 & 37.9 & 34.6 & 36.8 & 41.4 & 9.20 & 6.40 \\
    DocOwl2$^\dagger$ & 7B & $\sim$324 &
    57.5 & 53.2 & 45.2 & 44.7 & 47.5 & 45.2 & 48.9 & 3.94 & 0.07 \\
    TokenPacker$^\dagger$ & 7B & 144 &
    46.4 & 39.9 & 41.3 & 33.5 & 36.7 & 35.9 & 37.5 & 4.43 & \textbf{0.05} \\
    \midrule
    \multicolumn{12}{l}{\textbf{Ours}} \\
    \rowcolor{gray!10}
    \textbf{Doc-REFRAG} & 7B & $\sim$150 & \textbf{61.2} & \textbf{60.8} & \textbf{58.7} & \textbf{53.2} & \textbf{57.3} & \textbf{54.9} & \textbf{57.5} & \underline{2.57} & 0.07 \\
    \bottomrule
    \end{tabular}
    }
    \vspace{4pt}
   \caption{Comparison with general MLLMs and document-aware MLLMs on six benchmarks under \textbf{RAG settings}.
   The `$^\dagger$' indicates models that use visual token compression.
   `Token$^V$' denotes the average number of visual tokens per image.
   `\textbf{Bold}' indicates SOTA performance within the group and `\underline{Underline}' indicates second-best performance within the group.
   Accuracy (\%~$\uparrow$), TTFT, and TTIT are reported in seconds ($\downarrow$). \# denotes averaged metrics.}
   \vspace{-12pt}
    \label{tab:main_new}
\end{table*}

Next, we train on Doc-750K, a multi-image VDU dataset comprising documents split across multiple images, applying the same 20\% expansion per image. The resulting longer hybrid sequences challenge the decoder to perform cross-image reasoning under extended context.

Finally, we train on DocLongRAG, which simulates realistic RAG conditions. Unlike Doc-750K, DocLongRAG comprises images from heterogeneous sources exhibiting redundancy, semantic overlap, and contradiction. We reduce expansion to 10\% and additionally activate randomly selected irrelevant chunks, introducing a dual challenge: noisy heterogeneous context combined with sparse and imperfect token expansion, thereby better aligning the model with practical RAG deployment.

\paragraph{RL-Based Selector Training.}
The RL-based selector is a lightweight two-layer transformer that computes selection logits over all chunk embeddings. Given the question $\mathbf{t}$ and chunk sequence $\{\mathbf{c}_i^j\}$, the policy $\pi_\theta$ samples binary decisions $\delta_i^j \in \{0,1\}$ to determine whether each chunk should be expanded to its original $k$ tokens. The policy is constrained to select exactly $T = \lfloor p \cdot n \cdot L \rfloor$ chunks for expansion, where $p \in (0,1]$ controls the budget. We train a dedicated selector for each value of $k$ and demonstrate the influence in App.~\ref{appendix:chunk_size}. In our experiments, $p$ is uniformly sampled from $[0.1, 0.3]$ during training and fixed at 0.2 during inference, which is identified as optimal in App.~\ref{appendix:expansion_budget}.

We train the selector on DocLongRAG using a GRPO-style policy optimization framework with clipped PPO updates. The reward signal is the answer accuracy of the frozen decoder when processing the input sequence $\{\delta_i^j\}$ constructed from the selector's choices. As shown in Sec.~\ref{sec:selector_analysis}, this yields lower inference latency than ColQwen with competitive accuracy. Full training details, including trajectory sampling and advantage normalization, are provided in App.~\ref{appendix:selector_training}.

\section{Experiments}
\label{sec:experiments}
\subsection{Experimental Setup}
\paragraph{Dataset \& Benchmarks.}
For training, we perform single-image reconstruction with DocStruct4M, followed by multi-image CPT on OpenDocVQA, Doc-750K, and DocLongRAG. The selector is also trained on DocLongRAG. All test-benchmark images are excluded to prevent data leakage.
Benchmarks include: ChartQA~\cite{masry2022chartqa}, SlideVQA~\cite{tanaka2023slidevqa}, InfoVQA~\cite{mathew2022infographicvqa}, DUDE~\cite{van2023document}, ViDoRe-v2~\cite{mace2025vidore}, and MMDocIR~\cite{dong2025mmdocir}.

\vspace{-2pt}
\paragraph{Baselines.}
We compare Doc-REFRAG with MLLMs that support multi-image input, which excludes methods designed for single-image settings. First, we include general MLLMs: Qwen3-VL~\cite{bai2025qwen2}, MiniCPM-V-2.6~\cite{yao2024minicpm}, LLaVA-OneVision~\cite{li2024llava}, and InternVL3.5~\cite{chen2024internvl}. Second, we evaluate visual token compression variants: MQA-LLaVA~\cite{hu2024matryoshka}, EPIC~\cite{wen2025efficient}, and LLaVA-Mini~\cite{zhang2025llava}. Finally, we include models designed for document understanding: VDocGenerator~\cite{tanaka2025vdocrag}, Docopilot~\cite{duan2025docopilot}, TokenPacker~\cite{li2025tokenpacker}, and DocOwl2~\cite{docowl2}.

\vspace{-4pt}
\paragraph{Implementation details.}
Doc-REFRAG is built upon DocOwl2, adopting its encoder and decoder as the backbone. In single-image reconstruction, we fine-tune the linear vision-to-text alignment module in the encoder, along with a two-layer MLP projector (hidden dimension 4096, matching the decoder's output dimension), using a peak learning rate of $2\times10^{-4}$. In multi-image CPT, we employ LoRA (rank = 64)~\cite{hu2022lora} to fine-tune the decoder using AdamW ($1\times10^{-5}$ learning rate, 4\% linear warmup, cosine decay) with a batch size of 64. Both stages use cross-entropy loss over predicted tokens. The RL-based selector is trained with GRPO (clip ratio = 0.2). All stages train for 2 epochs. Experiments are conducted on 8 $\times$ NVIDIA RTX 3090 GPUs. Per-stage datasets and trainable parameters are detailed in App.~\ref{appendix:training_cost}.

\subsection{Main Result}
We evaluate Doc-REFRAG against eleven state-of-the-art MLLMs across six VDU benchmarks. To assess model robustness under realistic RAG settings, we \textbf{retrieve the top-20 images} for each question from its source benchmark with the state-of-the-art retriever Jina-Embeddings-V4~\cite{gunther2025jina}, and provide them as input to all MLLMs. These top-20 candidates cover 93.4\% of the gold images on average across the six benchmarks (Recall@20), indicating that retrieval recall is not the bottleneck in our evaluation. We present quantitative results in this section and defer qualitative examples to App.~\ref{appendix:qualitative}.

\paragraph{Quality analysis.}
As shown in Tab.~\ref{tab:main_new}, Doc-REFRAG with $k = 3$ outperforms both general MLLMs and specialized long-document MLLMs. This gain stems from two design aspects. First, these MLLMs are typically trained on short, coherent documents. This diverges from the noisy retrieval contexts in evaluation, which include redundant, conflicting, or irrelevant content that impairs reasoning. In contrast, Doc-REFRAG trains on DocLongRAG, which is explicitly designed to mimic such noisy retrieval contexts. Second, most MLLMs degrade with input length due to attention saturation, while Doc-REFRAG's question-guided compression dynamically reduces tokens while preserving task-critical information.

\vspace{-2pt}
\paragraph{Efficiency Analysis.}
\begin{table*}[t!]
    \centering
    \vspace{2pt}
    \resizebox{0.9\linewidth}{!}{
    \begin{tabular}{l|cccccc|cc}
        \toprule
        \textbf{Strategy} & \textbf{ChartQA} & \textbf{SlideVQA} & \textbf{InfoVQA} & \textbf{DUDE} & \textbf{Vidore} & \textbf{MMDocIR} & \textbf{Average} & \textbf{Latency} \\
        \midrule
        Random      & 39.4 & 42.5 & 48.3 & 33.2 & 38.6 & 30.8 & 38.8 & - \\
        Perplexity  & 46.5 & 48.3 & 47.2 & 39.6 & 40.4 & 40.3 & 43.7 & 20.7\\
        ColQwen     & \underline{56.6} & \underline{54.0} & \textbf{59.5} & \underline{49.8} & \underline{53.5} & \underline{52.3} & \underline{53.6} & \underline{9.3}\\
        \rowcolor{gray!10}
        \textbf{RL (Ours)} & \textbf{61.2} & \textbf{60.8} & \underline{58.7} & \textbf{53.2} & \textbf{57.3} & \textbf{54.9} & \textbf{57.5} & \textbf{3.7}\\
        \bottomrule
    \end{tabular}
    }
    \caption{Performance of different selection strategies for Doc-REFRAG. Accuracy (\%~$\uparrow$) and Latency in seconds ($\downarrow$) are measured on 8$\times$RTX 3090 GPUs.}
    \vspace{-6pt}
    \label{tab:strategy_comparison}
\end{table*}

We measure TTFT and TTIT latency under a batch size of 1 with 20 retrieved images per question (accuracy--latency trade-offs with more retrieved images are analyzed in App.~\ref{appendix:efficiency}). Within Doc-REFRAG, TTFT drops from 3.24\,s ($k{=}2$, 194 tokens) to 1.95\,s ($k{=}6$, 108 tokens), with $k{=}3$ chosen as the default at 2.57\,s for the best accuracy--efficiency trade-off (Tab.~\ref{tab:main_new}). Compared to Docopilot, the most accurate baseline, Doc-REFRAG incurs a slight TTFT increase but reduces TTIT by 2.18 s, accelerating end-to-end generation for long outputs. When compared to token-compressed MLLMs under a comparable visual token budget, Doc-REFRAG matches their TTFT while improving accuracy by over 18\%.

\vspace{-2pt}
\paragraph{Selector Analysis.}
\label{sec:selector_analysis}

Doc-REFRAG uses an RL selector to decide which compressed chunks to expand at decoding. We compare three strategies: (1) \textit{Random}, uniformly sampling chunks; (2) \textit{ColQwen}, selecting chunks with the highest question-chunk similarity, mirroring training; (3) \textit{Perplexity-based}, a heuristic ranking chunks by decoder perplexity given the question and chunk embeddings, with lower values indicating higher relevance.

As shown in Tab.~\ref{tab:strategy_comparison}, our RL-based selector outperforms random (38.8\%) and perplexity-based (43.7\%) strategies, achieving 57.5\% average accuracy. In comparison, ColQwen, a dedicated dense retriever for question-document relevance, achieves higher accuracy on InfoVQA, possibly because InfoVQA is part of its training data~\cite{faysse2024colpali}. However, ColQwen computes similarity scores over patches and aggregates them to infer chunk relevance, incurring a latency of 9.3 s. By contrast, our selector identifies relevant chunks in a single forward pass at 3.7 s.

\subsection{Analysis}
\paragraph{Can the chunk embedding handle dense text?}
Chunk embeddings risk losing textual details critical for RAG. We design a synthetic benchmark of A4 document images (595 $\times$ 842 pixels), synthesizing English passages via PyMuPDF with font sizes 10--20 and 500--6,000 characters per image.

\begin{figure}[!b]
    \centering
    \includegraphics[width=\columnwidth]{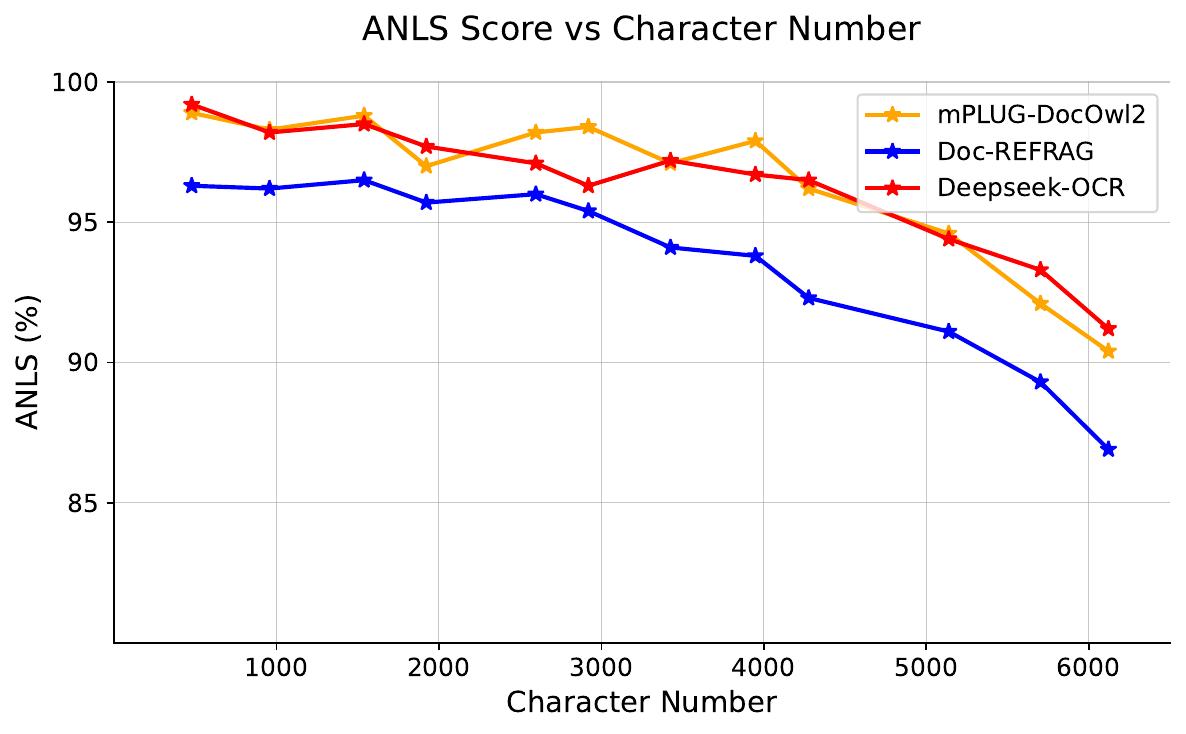}
    \vspace{-10pt}
    \caption{Parsing performance on A4 document image.}
    \vspace{-12pt}
    \label{fig:Parsing_performance}
\end{figure}

We evaluate all models by their ability to recover original text from images using Average Normalized Levenshtein Similarity (ANLS). As shown in Fig.~\ref{fig:Parsing_performance}, Doc-REFRAG (150 tokens/image) achieves ANLS above 95\% for documents up to 4,000 characters, matching DeepSeek-OCR (256 tokens)~\cite{wei2025deepseek} and DocOwl2 (324 tokens) despite its compact representation. Even at 6,000 characters, performance degrades only marginally, remaining within a 5\% absolute gap from the baselines. This demonstrates that chunk embeddings preserve textual semantics even under high lexical density.

\paragraph{Can our training tasks be beneficial?}

Tab.~\ref{tab:CPT_ablation} confirms that our training tasks are essential. Removing single-image reconstruction drops accuracy from 55.1\% to 19.7\%, showing that chunk embeddings lose semantic fidelity without reconstruction supervision. Replacing curriculum learning with randomly shuffled multi-image sequences reduces accuracy to 24.2\%, showing that mixed-token handling requires progressive exposure during training. These results validate that our training strategy enables effective compression while preserving textual alignment.

\begin{table}[!t]
    \centering
    \resizebox{\columnwidth}{!}{
    \begin{tabular}{l|ccc|c}
        \toprule
        \textbf{Variant} & \textbf{Vidore} & \textbf{MMDocIR} & \textbf{DUDE} & \textbf{Average} \\
        \midrule
        \rowcolor{gray!10}
        \textbf{Doc-REFRAG (Full)} & \textbf{53.2} & \textbf{57.3} & \textbf{54.9} & \textbf{55.1} \\
        w/o Single-image Recon. & 20.2 & 16.5 & 22.5 & 19.7 \\
        w/o Curriculum Learning & \underline{24.5} & \underline{22.8} & \underline{25.4} & \underline{24.2} \\
        w/o Both & 7.2 & 8.5 & 8.4 & 8.0 \\
        \bottomrule
    \end{tabular}
    }
    \caption{Ablation study of training tasks. Results are reported as accuracy ($\%\,\uparrow$) on three key benchmarks.}
    \vspace{-4pt}
    \label{tab:CPT_ablation}
\end{table}

\paragraph{Can our dataset improve the performance?}
\label{sec:dataset_ablation}
Conventional VDU datasets often lack the noise and redundancy characteristic of multi-image retrieval. Excluding DocLongRAG from training reduces Doc-REFRAG's average accuracy from 55.1\% to 48.9\%, underscoring its critical role in robust RAG adaptation. To assess generalizability, we fine-tune InternVL3.5, MiniCPM-V-2.6, and DocOwl2 on DocLongRAG, yielding accuracy improvements of 3.9\%, 4.3\%, and 3.9\%, respectively. This consistent cross-model gain validates DocLongRAG as the first long-context, challenging dataset for multimodal document RAG. A $2{\times}2$ factorial decomposition of architecture and data contributions is provided in App.~\ref{appendix:disentangle}. To validate that gains stem from the construction pipeline rather than mere data volume, we compare against a naive random baseline in App.~\ref{appendix:naive_baseline}.

\begin{table}[!htbp]
    \centering
    \resizebox{\columnwidth}{!}{
    \begin{tabular}{l|ccc|c}
        \toprule
        \textbf{Model} & \textbf{DUDE} & \textbf{Vidore} & \textbf{MMDocIR} & \textbf{Average} \\
        \midrule
        InternVL3.5 (orig.) & 45.8 & 42.1 & 51.5 & 46.5 \\
        InternVL3.5 (ft.)   & 47.1 & 50.4 & 53.8 & \underline{50.4} \\
        \midrule
        MiniCPM-V-2.6 (orig.) & 46.4 & 41.1 & 48.8 & 45.4 \\
        MiniCPM-V-2.6 (ft.)   & \underline{51.0} & 45.3 & 52.8 & 49.7 \\
        \midrule
        DocOwl2 (orig.)      & 44.7 & 47.5 & 45.2 & 45.8 \\
        DocOwl2 (ft.)        & 48.4 & \underline{51.6} & 49.1 & 49.7 \\
        \midrule
        Doc-REFRAG (w/o DocLongRAG)     & 48.5 & 43.7 & \underline{54.6} & 48.9 \\
        \rowcolor{gray!10}
        \textbf{Doc-REFRAG (Full)} & \textbf{53.2} & \textbf{57.3} & \textbf{54.9} & \textbf{55.1} \\
        \bottomrule
    \end{tabular}
    }
    \caption{Ablation study of DocLongRAG. General and document-aware MLLMs are compared in original (orig.) and fine-tuned (ft.) versions on DocLongRAG. ``Doc-REFRAG (w/o DocLongRAG)'' denotes our model trained without the DocLongRAG corpus. Results are reported as accuracy (\%~$\uparrow$).}
    \vspace{-2pt}
    \label{tab:dataset_ablation}
\end{table}

\paragraph{Can Doc-REFRAG help improve the reranker performance?}
Document reranking is commonly formulated as binary classification, where MLLMs sequentially assess the relevance of each candidate image to a given question via the logit of the ``Yes'' token. However, processing each image incurs high inference latency, prohibitive across many candidates.

Doc-REFRAG reduces per-document cost through efficient visual encoding. By globally compressing visual tokens and dynamically restoring question-relevant details, it enables faster relevance scoring without sacrificing accuracy. As shown in Tab.~\ref{tab:new_reranker}, Doc-REFRAG achieves higher reranking recall than dedicated rerankers on MMDocIR, yet incurs much lower per-document latency.
\begin{table}[t!]
    \centering
    \resizebox{\columnwidth}{!}{
    \begin{tabular}{l|ccc|c}
        \toprule
        \textbf{Model} & \textbf{MMDocIR} & \textbf{Vidore} & \textbf{DUDE} & \textbf{Latency} \\
        \midrule
        Jina-m0~\cite{gunther2025jina} & 81.1 & \underline{84.6} & \underline{86.2} & 13.6\\
        MM-R5~\cite{xu2025mm} & \underline{82.3} & 81.9 & 81.3 & 49.5\\
        MonoQwen2~\cite{MonoQwen} & 81.4 & 82.1 & 82.4 & \underline{12.5}\\
        \rowcolor{gray!10}
        \textbf{Doc-REFRAG} & \textbf{84.2} & \textbf{85.8} & \textbf{87.7} & \textbf{7.3} \\
        \bottomrule
    \end{tabular}
    }
    \vspace{-4pt}
    \caption{Performance comparison across rerankers. Results are reported as Recall@10 ($\uparrow$). Latency is measured in seconds ($\downarrow$).}
    \vspace{-8pt}
    \label{tab:new_reranker}
\end{table}

\paragraph{Can Doc-REFRAG surpass the reranker pipeline?}

We compare Doc-REFRAG against two standard retrieval-then-rerank baselines: (1) \textit{Direct Retrieval}, which feeds the top-5 candidates from the retriever Jina-Embeddings-V4 directly into Qwen3-VL-8B, and (2) \textit{Reranking Pipeline}, which uses MonoQwen2~\cite{MonoQwen} to rerank top-20 candidates to 5 before answering.

As shown in Tab.~\ref{tab:rerank_comparison}, the Direct Retrieval baseline suffers from evidence loss due to the performance ceiling of dual-tower retrievers, where limited top-$k$ candidates fail to encapsulate sufficient information. The Reranking Pipeline mitigates this loss, improving accuracy to 53.2\%, but incurs 15.4s latency from the LLM-based reranking stage ($\sim$0.63s/img). In contrast, Doc-REFRAG processes all top-20 candidates via question-guided compression, achieving 57.5\% accuracy in only 6.1s, resolving the trade-off between retrieval recall and inference efficiency in multimodal RAG.

\begin{table}
    \centering
    \resizebox{\columnwidth}{!}{
    \begin{tabular}{l|c|cc}
        \toprule
        \textbf{Pipeline Strategy} & \textbf{Selection Method} & \textbf{Latency} & \textbf{Acc.}  \\
        \midrule
        Direct Retrieval & Top-5 (No Reranker) & \textbf{4.2} & 46.6 \\
        Conventional Reranking & MonoQwen2 (Top-20$\rightarrow$5) & 15.4 & \underline{53.2} \\
        \rowcolor{gray!10}
        \textbf{Doc-REFRAG} & \textbf{Selective Encoding (Top-20)} & \underline{6.1} & \textbf{57.5} \\
        \bottomrule
    \end{tabular}
    }
    \vspace{-4pt}
    \caption{Comparison of different RAG pipelines. Results are reported as accuracy (\% $\uparrow$) averaged across six benchmarks, and total end-to-end time in seconds ($\downarrow$), including retrieval, (re-)ranking, and generation.}
    \label{tab:rerank_comparison}
    \vspace{-12pt}
\end{table}

\section{Conclusion}
We introduce Doc-REFRAG, an efficient decoding framework for multimodal document RAG. It employs question-guided compression and a three-stage training paradigm to compress visual tokens into chunks, selectively expanded via an RL-based selector. Experiments across six benchmarks show Doc-REFRAG outperforms eleven baselines, achieving state-of-the-art accuracy with lower latency. We further validate our training methodology, the role of DocLongRAG, and its transferability to document reranking. Our approach enables efficient large-context inference in RAG, especially in resource-constrained deployments.

\clearpage
\section*{Limitations}

While Doc-REFRAG advances efficient multimodal document RAG, three limitations remain. First, the chunk size $k$ is fixed as a training hyperparameter and cannot adapt dynamically at inference time. Although $k=3$ achieves the best overall accuracy--efficiency trade-off (App.~\ref{appendix:chunk_size}), performance degrades by approximately 5\% ANLS on documents exceeding 4,000 characters, indicating that adaptive granularity could further improve coverage in high-density layouts. Adapting $k$ at inference time, however, is not free. Each selector is trained at a fixed $k$, so supporting multiple chunk sizes requires an additional routing step that dispatches each input to the matching selector, whose latency works against the efficiency goal of the method. We leave a single selector that generalizes across chunk sizes to future work. Second, the RL-based selector is trained with answer accuracy as the sole reward signal, providing no fine-grained attribution of which individual chunks contributed to the correct answer. As shown in App.~\ref{appendix:qualitative}, 64\% of remaining errors concentrate on dense or structurally complex text regions, suggesting that richer reward shaping or chunk-level supervision could improve selector precision in these cases. Third, while our evaluation uses a real retriever over real documents, the training noise in DocLongRAG is constructed, and no single construction pipeline covers every deployment scenario. DocLongRAG is also limited to English documents. Although the construction pipeline is language-agnostic, dense CJK text packs more information per patch and may favor a smaller chunk size. Finally, due to the high computational cost of training large multimodal models, we report results from a single training run, which is standard practice in the multimodal document understanding community. All baseline comparisons follow the same single-run protocol.

\section*{Ethics Statement}

DocLongRAG is constructed from publicly available datasets and document corpora. During dataset construction, we apply a four-stage filtering pipeline that removes context-independent questions, online-search-dependent queries, unanswerable instances, and ethically problematic content including personally identifiable information, offensive language, and sensitive imagery (Sec.~\ref{sec:dataset}). The synthetic document images used in our dense-text evaluation are generated programmatically and contain no real user data.

Doc-REFRAG is designed as a document understanding tool and does not store or transmit user documents beyond the scope of inference. Although grounding responses in retrieved documents reduces hallucination risk relative to purely parametric models, the system can still produce errors in documents with dense, overlapping, or contradictory evidence. Deployment in high-stakes decision-making contexts requires human oversight and should not rely solely on model outputs.

\section*{Acknowledgments}
This work was supported by the National Natural Science Foundation of China under Grant No. U24A20326. We thank the anonymous reviewers and action editors for their constructive feedback.

\bibliography{main}

\clearpage
\appendix
\section{Differentiated Contributions Over Related Compression Methods}
\label{appendix:differentiation}

We provide a structured comparison of Doc-REFRAG against the most closely related visual token compression methods across six dimensions in Tab.~\ref{tab:differentiation}. The comparison covers methods that appear as baselines in our main evaluation (Tab.~\ref{tab:main_new}), DocOwl2 (our base model), and three additional query-conditioned compression methods (SparseVILA, IVTP, LLaMA-VID) that are most architecturally similar to ours.

\begin{table*}[ht]
\centering
\resizebox{\linewidth}{!}{%
\begin{tabular}{lcccccc}
\toprule
\textbf{Method} &
\textbf{Query-conditioned} &
\textbf{Dynamic Budget} &
\textbf{RL-trained Selector} &
\textbf{RAG Noise Robust.} &
\textbf{Learned Restoration} &
\textbf{Multi-image RAG} \\
\midrule
MQT-LLaVA~\cite{hu2024matryoshka}   & \xmark & \cmark & \xmark & \xmark & \xmark & \xmark \\
EPIC~\cite{wen2025efficient}         & \xmark & \xmark & \xmark & \xmark & \xmark & \xmark \\
LLaVA-Mini~\cite{zhang2025llava}     & \xmark & \xmark & \xmark & \xmark & \xmark & \xmark \\
TokenPacker~\cite{li2025tokenpacker} & \xmark & \xmark & \xmark & \xmark & \xmark & \xmark \\
DocOwl2~\cite{docowl2}               & \xmark & \xmark & \xmark & \xmark & \xmark & \cmark \\
\midrule
SparseVILA~\cite{Khaki_2025} & \cmark & \xmark & \xmark & \xmark & \xmark & \xmark \\
IVTP~\cite{ye2024ivtp}               & \cmark & \xmark & \xmark & \xmark & \xmark & \xmark \\
LLaMA-VID~\cite{li2024llama}      & \cmark & \cmark & \xmark & \xmark & \xmark & \xmark \\
\midrule
\rowcolor{gray!10}
\textbf{Doc-REFRAG (Ours)}           & \cmark & \cmark & \cmark & \cmark & \cmark & \cmark \\
\bottomrule
\end{tabular}%
}
\vspace{4pt}
\caption{Comparison of Doc-REFRAG with related visual token compression methods across six design dimensions. \textbf{Query-conditioned}: selection conditioned on the input question. \textbf{Learned Restoration}: selective restoration of compressed tokens via a learned policy (as opposed to training-free heuristic retrieval).}
\label{tab:differentiation}
\end{table*}

We elaborate on the dimensions where Doc-REFRAG introduces substantive novelty.

\paragraph{Query-conditioned compression.}
Methods such as SparseVILA, IVTP, and LLaMA-VID condition token selection or compression on the input query. However, their query-conditioning mechanisms are heuristic: SparseVILA scores visual tokens against the query using attention at decode time (training-free); IVTP aggregates text instructions into a CLS token to guide pruning via attention rollout; LLaMA-VID generates a query-dependent context token alongside fixed content tokens. None of these methods \emph{learn} a selection policy through end-to-end optimization against a downstream task reward. Doc-REFRAG trains its selector via GRPO, directly optimizing chunk selection for answer accuracy---making the selection policy a first-class learned component rather than a post-hoc heuristic.

\paragraph{Dynamic budget with RL training.}
Prior methods either fix the compression ratio at design time (IVTP, EPIC, LLaVA-Mini) or apply query-independent dynamic sparsity (SparseVILA). Doc-REFRAG adopts a dynamic budget $p \in [0.1, 0.3]$ sampled during training, enabling the model to learn budget-robust selection policies via GRPO with clipped PPO updates, using the frozen decoder's answer accuracy as the reward signal---a training paradigm not previously applied to visual token compression.

\paragraph{RAG noise robustness.}
All existing compression methods are developed and evaluated on coherent single-source documents or video frames. Doc-REFRAG is trained on DocLongRAG, which explicitly simulates realistic retrieval noise: retrieved images are heterogeneous, may be redundant or contradictory, and include hard negatives interleaved with relevant content. This distributional alignment between training and deployment is a key factor behind Doc-REFRAG's performance advantage in RAG settings (Tab.~\ref{tab:main_new}).

\paragraph{Learned restoration vs.\ heuristic retrieval vs.\ condensation.}
SparseVILA retains the full KV cache at prefill and selectively retrieves tokens per decoding step---a form of training-free heuristic restoration, not a learned policy. TokenPacker and DocOwl2 implement coarse-to-fine \emph{condensation}: progressively merging tokens into compact representations (irreversible and question-agnostic). Doc-REFRAG instead implements \emph{learned restoration}: it first compresses all tokens into chunk embeddings via an MLP projector, then an RL-trained selector dynamically identifies and restores original fine-grained tokens for question-relevant chunks. The restoration decision is learned end-to-end against task reward, not computed by a heuristic. This makes Doc-REFRAG's approach fundamentally distinct from both the condensation paradigm (TokenPacker, DocOwl2) and the training-free retrieval paradigm (SparseVILA).

\section{Source Datasets in DocLongRAG Corpus}
\label{appendix:source_datasets}
DocLongRAG aggregates English data from the following 15 open-source datasets:~\cite{van2023document,masry2022chartqa,mathew2021docvqa,ye2023mplug,mathew2022infographicvqa,tanaka2021visualmrc,tanaka2023slidevqa,zhang2023mpmqa,laurenccon2024building,liu2024mmdu,chen2023longlora,yang2023longqlora,zhang2024longcite,bai2024longalign,zhang2024longreward}.

\section{Evaluation of Hard Negative Relevance}
\label{appendix:hard_negative}

\paragraph{Limitations about the annotation.}
Here, we present an example from the benchmarks to illustrate their limitations regarding the relevance annotation, which emphasizes the importance of negative samples exclusion.

\begin{figure*}[t]
    \centering
    \begin{subfigure}[b]{0.49\textwidth}
        \centering
        \includegraphics[width=\textwidth]{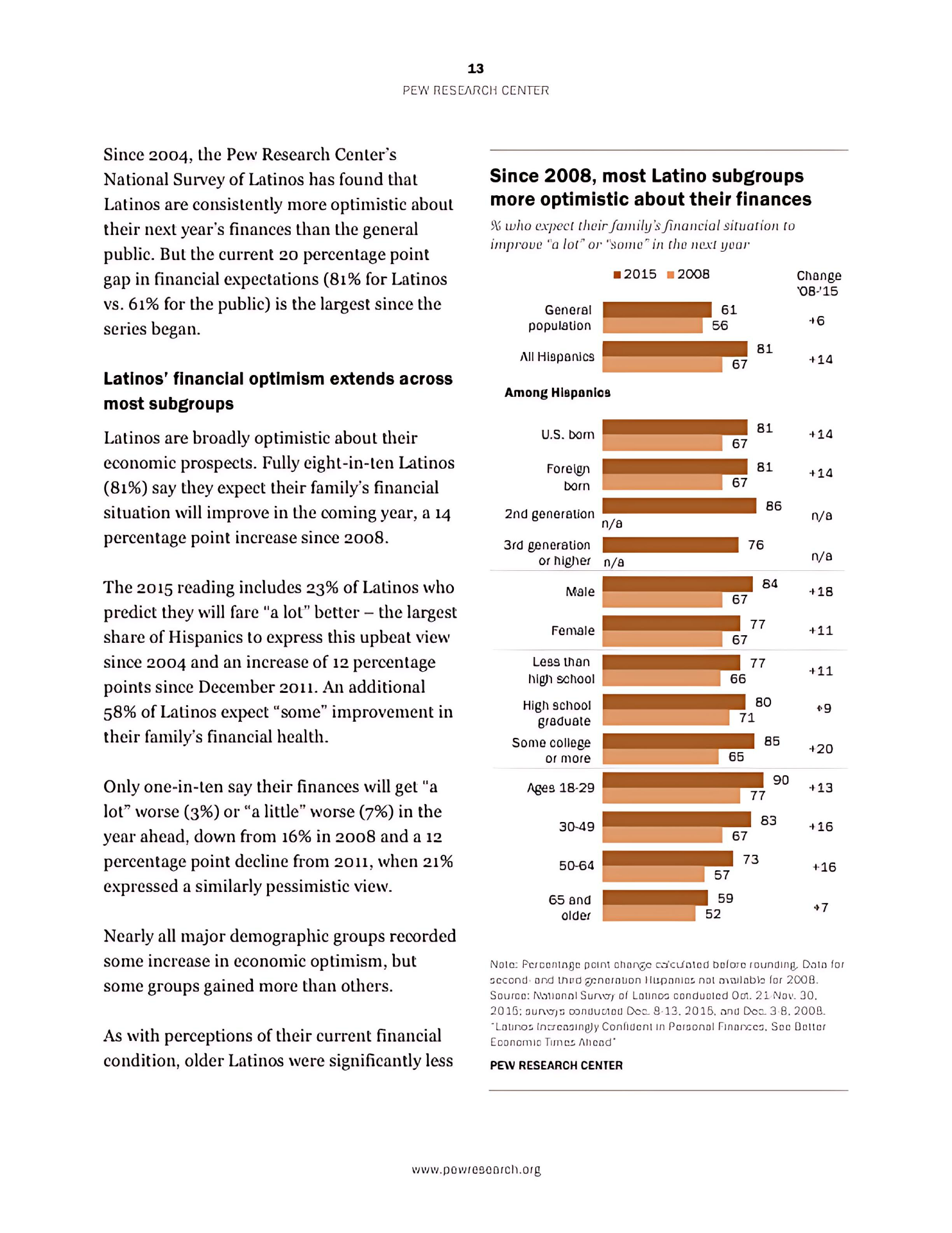}
        \caption{Ground Truth for Query}
        \label{fig:example1}
    \end{subfigure}
    \hfill
    \begin{subfigure}[b]{0.49\textwidth}
        \centering
        \includegraphics[width=\textwidth]{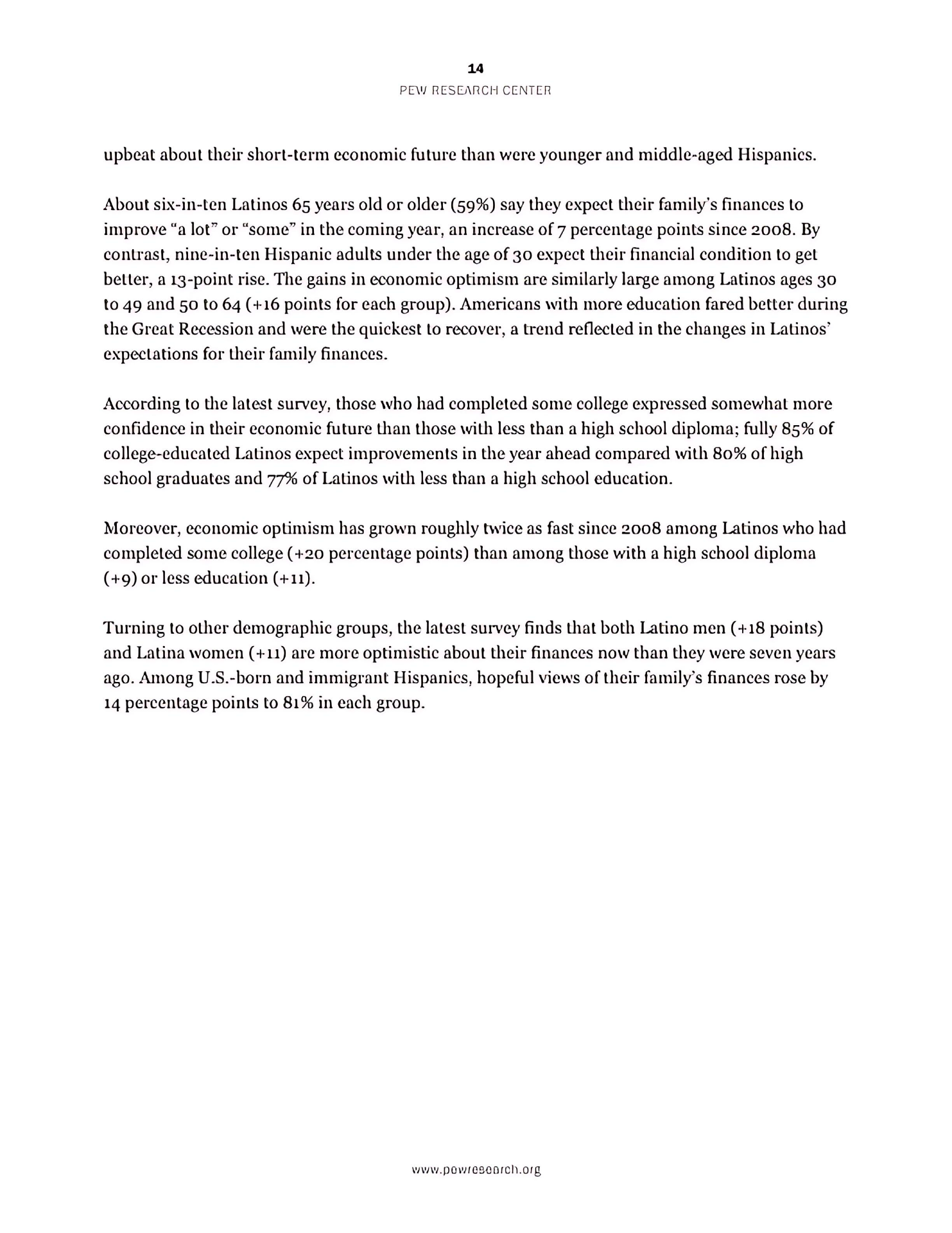}
        \caption{Another Proper Page for Query}
        \label{fig:example2}
    \end{subfigure}
    \caption{Example for Benchmark.}
    \label{fig:examples}
\end{figure*}

When the query is \texttt{[From this report, which subgroup among Hispanics has gained most confidence from 2008 to 2015?]}, the original benchmark MMDocIR designates Fig.~\ref{fig:example1} as the exclusive retrieval target, as it provides a bar chart comparing financial optimism across various Latino subgroups from 2008 to 2015. The 08-15 column highlights the percentage-point increase in optimism for each group, with the ``Some college or more'' subgroup exhibiting the largest change (+20 percentage points), reflecting the greatest gain in confidence during this period.

However, Fig.~\ref{fig:example2} also constitutes a valid retrieval target. By explicitly referencing 2008 and noting that this year was ``7 years ago,'' the document confirms that the data summarizes trends from 2008 to 2015. Furthermore, it specifies that economic optimism among Latinos with some college education increased by +20 percentage points---roughly twice the growth observed among those with only a high school diploma (+9) or less education (+11). This reinforces that the ``Some college or more'' subgroup experienced the most substantial rise in confidence. Thus, the pair from the raw benchmark \textbf{overlooks an equally relevant candidate}, which hinders the fair evaluation for retrieval systems.

\paragraph{Human validation about the hard negative relevance.}
To rigorously validate the semantic relevance of non-annotated images and inform the design of hard negative sampling in DocLongRAG, we conduct a controlled human evaluation study.

We randomly select 500 questions from the curated corpus $\mathcal{C}$, each associated with a set of document images retrieved by a document-aware retriever. For each query, we examine the top-20 retrieved images that do not belong to the ground-truth relevant set $\mathcal{D}$. Specifically, we analyze images at retrieval ranks 1 through 20 to quantify how semantic relevance decays with decreasing rank.

Two expert annotators participate in the evaluation. Given a query $q$, its ground-truth answer $a$, and a candidate image $d$, each annotator rates the degree to which $d$ contains content that supports answering $q$. Relevance is assessed on a 5-point Likert scale defined as follows:

\begin{itemize}
    \item \textbf{5}: The image contains explicit and sufficient evidence to answer the question.
    \item \textbf{4}: The image provides partial or indirect evidence that strongly supports the answer.
    \item \textbf{3}: The image is topically related but lacks direct evidential value.
    \item \textbf{2}: The image shares superficial similarity (e.g., same document type or layout) but is not helpful for answering.
    \item \textbf{1}: The image is clearly irrelevant.
\end{itemize}

Annotators work independently, and inter-annotator agreement is measured using Krippendorff's alpha, yielding a value of 0.78, which indicates substantial agreement.

The average relevance scores exhibit a clear decay trend across retrieval ranks: images at rank 1, 2, and 3 receive average scores of 4.62, 4.31, and 3.94, respectively, indicating that they frequently contain meaningful supporting evidence. At rank 4 and onward, the average score remains lower than 3.27, suggesting limited relevance. This pattern confirms that images beyond rank 3 are substantially less likely to provide evidential support for the correct answer. Even so, the relevance at rank 4 and beyond is not zero, so a small fraction of hard negatives may still contain answer evidence. This does not corrupt the training objective: the annotated relevant image is always retained in the sequence by construction, so the supervision remains well-defined, and such cases act as additional retrieval noise rather than false supervision.

\section{Placing Meta-Chunk at Different Positions}
\label{appendix:metachunkposition}

We explore three different strategies for combining meta-chunk and hard negatives, which are represented by the following descriptions:

\begin{enumerate}[itemsep=0pt]
    \item \textbf{Head}: Placing the meta-chunk at the beginning of the retrieved hard negatives.
    \item \textbf{Tail}: Placing the meta-chunk at the end of the retrieved hard negatives.
    \item \textbf{Random}: Randomly inserting the meta-chunk within the retrieved hard negatives.
\end{enumerate}

Tab.~\ref{tab:meta_position_table} shows that placing the meta-chunk at the beginning (Head) yields better performance. Placing relevant images before hard negatives establishes longer dependencies between the question-relevant content and the decoder, which enhances overall reasoning effectiveness.

\begin{table*}[ht]
    \centering
    \resizebox{0.7\linewidth}{!}{%
    \begin{tabular}{l|cccccc|c}
        \toprule
        \textbf{Model} & \textbf{ChartQA} & \textbf{SlideVQA} & \textbf{InfoVQA} & \textbf{DUDE} & \textbf{Vidore} & \textbf{MMDocIR} & \textbf{\#Acc.} \\
        \midrule
        Head   & 61.2 & 60.8 & 58.7 & 53.2 & 57.3 & 54.9 & 57.5 \\
        Tail   & 50.1 & 52.6 & 50.7 & 40.8 & 40.7 & 44.9 & 46.6 \\
        Random & 58.5 & 54.1 & 53.5 & 45.4 & 42.8 & 53.0 & 51.2 \\
        \bottomrule
    \end{tabular}%
    }
    \vspace{4pt}
    \caption{Performance Comparison of Different Insertion Strategies.}
    \label{tab:meta_position_table}
\end{table*}

\paragraph{Training-time ordering and inference-time robustness.}
We investigate whether the head-position advantage (Tab.~\ref{tab:meta_position_table}) introduces a positional bias at inference time, causing the decoder to exploit token position rather than semantic content.

We emphasize that the meta-chunk ordering strategy studied here is a \emph{training-time} design choice for CPT, while at inference time the RL-based selector determines which chunks are expanded based purely on \emph{content relevance} to the question---not position. Specifically, the selector operates on chunk embeddings from all retrieved images jointly and assigns expansion decisions $\delta_i^j \in \{0,1\}$ via a learned policy $\pi_\theta$ trained with GRPO rewards on answer accuracy. The input to the selector is a set of chunk-level semantic embeddings; the reward signal is answer accuracy, which is invariant to the physical ordering of images in the context. As a consequence, the training objective provides no gradient signal that encourages position-dependent selection behavior.

This design ensures that the head-position benefit observed in Tab.~\ref{tab:meta_position_table} stems from improved long-range dependency modeling during training---not from a spurious positional shortcut exploited at inference. The meta-chunk always appears at the head of the \emph{training} sequence, but at inference time the selector re-evaluates every chunk independently and can assign high expansion probability to any image regardless of its position in the retrieved list.

\section{System Components: Architecture, Training, and Inference}
\label{appendix:system_overview}

To clarify the roles of Doc-REFRAG's components across three conceptually distinct phases, we provide a consolidated overview in Tab.~\ref{tab:system_overview}. Each component is associated with a specific phase (Architecture, Training, or Inference) and a brief description of its role, with exact hyperparameters for cross-referencing with Secs.~\ref{sec:dataset}--\ref{sec:experiments}.

\begin{table*}[ht]
\centering
\resizebox{\linewidth}{!}{%
\begin{tabular}{llll}
\toprule
\textbf{Component} & \textbf{Phase} & \textbf{Description} & \textbf{Details} \\
\midrule
Vision Encoder & Architecture & Encodes each image into 324 visual tokens & Inherited from DocOwl2; frozen after Stage 1 \\
MLP Projector ($\phi$) & Architecture & Maps $k$ visual tokens to one chunk embedding $\mathbf{c}_i^j \in \mathbb{R}^d$ & 2-layer MLP, hidden dim 4096; trained in Stage 1 \\
Decoder (LLM) & Architecture & Autoregressively generates the answer from the reconstructed token sequence & 7B decoder with LoRA fine-tuned in Stage 2 \\
RL Selector ($\pi_\theta$) & Architecture & Predicts binary expansion decisions $\delta_i^j$ for each chunk & 2-layer Transformer, hidden dim 256, 8 heads; $<$1M params \\
\midrule
Stage 1: Single-Image Reconstruction & Training & Aligns MLP projector with decoder embedding space & DocStruct4M; decoder frozen; lr $2\!\times\!10^{-4}$ \\
Stage 2: Multi-image CPT & Training & Scales to multi-image hybrid sequences with pseudo-label expansion & OpenDocVQA + Doc-750K + DocLongRAG; LoRA rank 64; lr $1\!\times\!10^{-5}$ \\
Stage 3: RL Selector Training & Training & Trains selector to maximize answer accuracy via GRPO & DocLongRAG; GRPO + clipped PPO (clip 0.2); reward = decoder accuracy \\
\midrule
Chunk Encoding & Inference & Compresses each image to $L = m/k$ chunk embeddings & $m=324$ tokens/image, $k=3$, $L=108$ chunks/image \\
Selector Forward Pass & Inference & Single forward pass to predict which chunks to expand; budget $p=0.2$ & Expands top $T = \lfloor 0.2 \cdot n \cdot L \rfloor$ chunks; latency 3.7 s vs.\ 9.3 s (ColQwen) \\
Chunk Expansion & Inference & Selected chunks restored to original $k$ visual tokens before decoding & Enables fine-grained detail recovery for question-relevant regions \\
\bottomrule
\end{tabular}%
}
\vspace{4pt}
\caption{System overview of Doc-REFRAG, organized by architectural components, training stages, and inference-time operations.}
\label{tab:system_overview}
\end{table*}

\section{Geometric Mapping Between Coarse and Fine Patch Grids}
\label{appendix:geometric_mapping}

To utilize question-to-patch similarity scores from ColQwen as pseudo-labels in CPT, we establish a deterministic geometric mapping between the coarse spatial grid used by our encoder and the fine-grained patch grid employed by ColQwen.

Both models discretize the input images into spatial patches using normalized image coordinates, where each pixel location is represented as $(x, y) \in [0, 1]^2$. Let $G_c$ denote the coarse grid used by CPT, with patch size $s_c = \frac{1}{H_c} \times \frac{1}{W_c}$, where $H_c$ and $W_c$ are the number of horizontal and vertical divisions. Similarly, let $G_f$ denote the fine grid used by ColQwen, with patch size $s_f = \frac{1}{H_f} \times \frac{1}{W_f}$, where $H_f > H_c$ and $W_f > W_c$.

For any coarse patch $p_c \in G_c$, its spatial extent is defined by the bounding box:
\[
p_c = \left[ i \cdot s_c,\ (i+1) \cdot s_c \right] \times \left[ j \cdot s_c,\ (j+1) \cdot s_c \right],
\]
where $(i, j)$ indexes the patch in the coarse grid.

Each such coarse patch $p_c$ covers a subset of fine patches in $G_f$. Specifically, the set of fine patches contained within $p_c$ is given by:
\[
\mathcal{F}(p_c) = \left\{ p_f \in G_f \mid \text{bbox}(p_f) \subseteq \text{bbox}(p_c) \right\}.
\]

Due to the alignment of coordinate systems and uniform sampling, this containment relationship is fully deterministic. We compute $\mathcal{F}(p_c)$ by iterating over all fine patches and checking whether their bounding boxes lie entirely within $p_c$. In practice, this can be implemented via integer arithmetic based on the ratio $r = \frac{H_f}{H_c} = \frac{W_f}{W_c}$, which must be an integer for perfect alignment.

This mapping allows us to aggregate the similarity scores from the fine patches in $\mathcal{F}(p_c)$ to generate a pseudo-label $\delta_i^j$ for the importance of coarse chunk $i$ in question $j$. The aggregation is performed as:
\[
\delta_i^j = \frac{1}{|\mathcal{F}(p_c)|} \sum_{p_f \in \mathcal{F}(p_c)} \text{sim}(q_j, p_f),
\]
where $\text{sim}(q, p_f)$ denotes the question-to-patch similarity score provided by ColQwen.

This deterministic and spatially aligned mapping ensures that the pseudo-labels are geometrically consistent and preserve local structure, enabling effective supervision of chunk selection during training.

\section{Robustness Analysis of Pseudo-Labels}
\label{appendix:pseudo_label_robustness}

Since ground-truth chunk-level relevance labels are unavailable in our setting, we conduct a human evaluation study to assess the validity of the binary pseudo-labels $\delta_i^j \in \{0, 1\}$ generated in App.~\ref{appendix:geometric_mapping}.

We randomly sample 200 question--document pairs from DocLongRAG. For each document, we compute $\delta_i^j$ from the aggregated ColQwen similarity scores. Two expert annotators, familiar with document understanding tasks, independently label each chunk as relevant or irrelevant based on whether it contains content that directly or indirectly supports answering the given question. The inter-annotator agreement (Cohen's $\kappa$) is 0.79, indicating substantial agreement.

We compare the pseudo-labels against the human-annotated labels. The pseudo-labels achieve an accuracy of 86.4\%, with a precision of 84.1\%, recall of 80.3\%, and F1 score of 0.82. These results demonstrate that the pseudo-labels align well with human judgments of chunk relevance, providing strong empirical justification for their use as supervision signals in training.

\section{Analysis of Chunk Size $k$}
\label{appendix:chunk_size}

In this section, we provide a detailed comparison of the performance and efficiency of Doc-REFRAG under different chunk sizes $k \in \{2, 3, 6\}$. As discussed in the main paper, the hyperparameter $k$ determines the granularity of visual token compression, where the original visual tokens of each image are partitioned into non-overlapping chunks of size $k$.

Tab.~\ref{tab:main_supple} presents the quantitative results across six document-level benchmarks. Our experiments reveal that $k=3$ achieves the optimal balance between semantic fidelity and computational efficiency, yielding the highest average accuracy of 57.5\%. Specifically, compared to $k=2$, the $k=3$ configuration improves average accuracy by 2.0\% while maintaining comparable inference latency. This suggests that a slightly coarser compression ($k=3$) effectively filters out redundant visual noise without sacrificing critical details, whereas $k=2$ may retain excessive irrelevant tokens that hinder the selector's decision-making.

On the other hand, increasing the chunk size to $k=6$ significantly reduces the number of visual tokens (from $\sim$194 to $\sim$108 per image) and lowers the Time-To-First-Token (TTFT) latency from 2.57s to 1.95s. However, this comes at the cost of a noticeable performance drop, with average accuracy decreasing by 5.2\% compared to $k=3$. The degradation is particularly evident in data-intensive tasks like ChartQA and InfoVQA, indicating that overly aggressive compression ($k=6$) compromises the preservation of fine-grained textual and structural details essential for complex reasoning.

Based on these findings, we adopt $k=3$ as the default setting for Doc-REFRAG in our main evaluations, as it delivers state-of-the-art accuracy while ensuring efficient inference suitable for practical RAG applications.

\begin{table*}[ht]
    \centering
    \resizebox{\linewidth}{!}{
    \begin{tabular}{lcc|cccccc|ccc}
    \toprule
     \textbf{Model} & \textbf{Size} & \textbf{Token$^V$} &
     \textbf{ChartQA} &
     \textbf{SlideVQA} &
     \textbf{InfoVQA} &
     \textbf{DUDE} &
     \textbf{Vidore} &
     \textbf{MMDocIR} &
     \textbf{\#Acc.} & \textbf{\#TTFT} & \textbf{\#TTIT} \\
    \midrule
    Doc-REFRAG$_{k=2}$ & 7B & $\sim$194 &
    58.6 & 56.2 & \underline{56.4} & \underline{51.9} & \underline{55.1} & \underline{53.7} & \underline{55.5} & 3.24 & 0.07 \\
    \rowcolor{gray!10}
    \textbf{Doc-REFRAG$_{k=3}$} & \textbf{7B} & $\sim$\textbf{150} &
    \textbf{61.2} & \textbf{60.8} & \textbf{58.7} & \textbf{53.2} & \textbf{57.3} & \textbf{54.9} & \textbf{57.5} & \textbf{2.57} & \textbf{0.07} \\
    Doc-REFRAG$_{k=6}$ & 7B & $\sim$108 &
    53.2 & 54.8 & 53.9 & 48.5 & 52.4 & 52.2 & 52.3 & \underline{1.95} & 0.07 \\
    \bottomrule
    \end{tabular}
    }
    \vspace{4pt}
    \caption{Comparison of Doc-REFRAG performance with different chunk sizes $k$. $k=3$ achieves the best accuracy--efficiency trade-off. TTFT denotes Time-To-First-Token.}
    \label{tab:main_supple}
\end{table*}

\section{Optimal Chunk Expansion Budget}
\label{appendix:expansion_budget}

The budget parameter $p \in (0, 1]$ controls the number of chunks selected for expansion, defined as $T = \lfloor p \cdot n \cdot L \rfloor$, where $n \cdot L$ is the total number of chunks. During training, $p$ is uniformly sampled from the interval $[0.1, 0.3]$ to encourage exploration and prevent overfitting to a fixed budget. To determine the optimal value of $p$ during inference, we conduct a hyperparameter search on the six benchmarks.

Specifically, we primarily fix the $p$ to values in the list $\{0.1, 0.15, 0.2, 0.25, 0.3\}$ and measure the answer accuracy of Doc-REFRAG when processing the expanded input sequence generated by the selector. As shown in Tab.~\ref{tab:p_search}, $p = 0.2$ achieves the highest accuracy (57.5\%) while maintaining low inference latency. Values below 0.2 provide insufficient context, compromising accuracy, whereas values above 0.2 incur higher computational costs without yielding significant improvements. This demonstrates that $p = 0.2$ achieves an optimal balance between performance and efficiency.

\begin{table}[ht]
\centering
\resizebox{0.7\columnwidth}{!}{%
    \begin{tabular}{c|c|c|c}
    \toprule
    $p$ & Accuracy (\%) & TTFT (s) & TTIT (s)\\
    \midrule
    0.1 & 51.6 & 2.19 & 0.07 \\
    0.15 & 54.2 & 2.34 & 0.07 \\
    0.2 & \textbf{57.5} & 2.57 & 0.07 \\
    0.25 & \underline{56.4} & 2.62 & 0.07 \\
    0.3 & 53.7 & 2.84 & 0.07 \\
    \bottomrule
    \end{tabular}
}
\vspace{4pt}
\caption{Answer accuracy under different chunk expansion budgets $p$.}
\label{tab:p_search}
\end{table}

\paragraph{Comparison with ColQwen across the full budget range.}
Tab.~\ref{tab:p_search} also enables a direct accuracy--latency comparison against ColQwen, which serves as an upper-bound selector oracle in our framework (Sec.~\ref{sec:selector_analysis}). ColQwen achieves 53.6\% average accuracy at a latency of 9.3\,s. Our RL-based selector at $p=0.2$ surpasses ColQwen with 57.5\% accuracy while requiring only 3.7\,s of selection latency---a $2.5\times$ reduction. At $p=0.25$ the selector reaches 56.4\% in 2.62\,s, and at $p=0.3$ it reaches 53.7\% in 2.84\,s, all remaining well below ColQwen's latency budget. Conversely, even at $p=0.1$ (2.19\,s, 51.6\%), the selector is substantially faster than ColQwen while sacrificing only 2.0 percentage points of accuracy. Across the entire search range $p \in \{0.1, \ldots, 0.3\}$, the RL selector consistently occupies a more favorable accuracy--latency operating point than ColQwen, confirming that replacing the dense retriever with a lightweight learned selector does not require trading accuracy for speed.

\section{Training Procedure for the Selector}
\label{appendix:selector_training}

We train the chunk selection policy using the GRPO framework. The full training procedure is outlined below.

\paragraph{Policy Network Architecture.}
The selector is a lightweight two-layer transformer encoder with hidden dimension 256 and 8 attention heads. Given a question embedding $\mathbf{t}$ and a sequence of chunk embeddings $\mathbf{c}^j_i$, the policy network computes logits over each chunk:
$$
\delta^j_i = \mathbf{W}_o \cdot \text{MLP}\left( \text{Transformer}(\mathbf{t}, \mathbf{c}_i^j) \right),
$$
where $\mathbf{W}_o$ is a learnable output weight matrix. The policy $\pi_\theta$ samples binary decisions $\delta_i^j \in \{0, 1\}$ independently for each chunk.

\paragraph{Group-Wise Relative Advantage Estimation.}
For each question-document pair, we sample $G=8$ chunk selection policies $\{\delta^{(1)}, \delta^{(2)}, \ldots, \delta^{(G)}\}$ from $\pi_\theta$. For each policy, we reconstruct the input sequence and compute the reward $r^{(g)}$ using the fixed decoder from the CPT stage. The relative advantage is defined as:
$$
A_i = r^{(i)} - \frac{1}{G} \sum_{g=1}^G r^{(g)},
$$
which measures the performance of the $i$-th policy relative to the group mean.

\paragraph{Policy Update with Clip and KL Regularization.}
The policy is updated using the following objective:
\begin{align}
\mathcal{L} = \ & \mathbb{E}_{\delta \sim \pi_\theta} \Big[
    \min\Big( \rho(\delta) A(\delta), \nonumber \\
& \quad\quad \text{clip}(\rho(\delta), 1{-}\epsilon, 1{+}\epsilon)\, A(\delta) \Big)
\Big] \nonumber \\
& - \beta \cdot D_{\mathrm{KL}}\big( \pi_\theta \,\|\, \pi_{\text{ref}} \big),
\end{align}
where $\rho(\delta) = \pi_\theta(\delta) / \pi_{\text{ref}}(\delta)$ is the importance sampling ratio, $\epsilon=0.2$, $\beta=0.1$, and $\pi_{\text{ref}}$ is the reference model initialized with SFT weights.

\section{Training Cost Analysis}
\label{appendix:training_cost}

The four components are not trained separately at full cost: only Stage 2 carries a non-trivial cost, while Stages 1 and 3 are lightweight. We provide a stage-by-stage breakdown of training configuration across the three stages in Tab.~\ref{tab:training_cost}. All experiments are conducted on 8$\times$ NVIDIA RTX 3090 GPUs (24 GB VRAM each). The base model is DocOwl2 (7B parameters).

\begin{table}[ht]
\centering
\resizebox{\columnwidth}{!}{%
\begin{tabular}{lccc}
\toprule
\textbf{Stage} & \textbf{Dataset} & \textbf{Trainable Parameters} & \textbf{Epochs} \\
\midrule
Stage 1: Single-Image Recon. & DocStruct4M & Encoder align.\ + MLP projector & 2 \\
Stage 2: Multi-image CPT & OpenDocVQA + Doc-750K + DocLongRAG & LoRA (rank 64) on 7B decoder & 2 \\
Stage 3: RL Selector & DocLongRAG & Selector ($<$1M params) & 2 \\
\bottomrule
\end{tabular}%
}
\vspace{4pt}
\caption{Training cost summary for Doc-REFRAG, broken down by stage, dataset, and trainable parameters.}
\label{tab:training_cost}
\end{table}

\paragraph{Stage 1} is computationally lightweight: the decoder is fully frozen and only the linear alignment module in the vision encoder and the two-layer MLP projector are updated. The small number of trainable parameters and single-image inputs (DocStruct4M~\cite{docowl15}) make this stage converge quickly.

\paragraph{Stage 2} constitutes the dominant cost. Training on the combined corpus of OpenDocVQA~\cite{tanaka2025vdocrag}, Doc-750K, and DocLongRAG---totaling over 1.1M question--image pairs---requires processing substantially longer multi-image hybrid sequences. Despite using LoRA (rank=64) to reduce the number of trainable parameters in the 7B decoder, the dataset scale and extended context lengths make this stage the bottleneck of the overall pipeline.

\paragraph{Stage 3} is again lightweight because the selector itself is a tiny two-layer transformer ($<$1M parameters); the 7B decoder is frozen and used only for reward evaluation on DocLongRAG with GRPO group size $G=8$. The modest parameter count and single-dataset scope ensure this stage adds minimal overhead relative to Stage 2.

Overall, the use of LoRA in Stage 2 and a lightweight selector in Stage 3 keeps the total training footprint manageable on commodity RTX 3090 hardware, making Doc-REFRAG accessible without specialized data-center resources.

\section{Qualitative Analysis}
\label{appendix:qualitative}

\begin{figure}[ht]
\centering
\includegraphics[width=0.6\columnwidth]{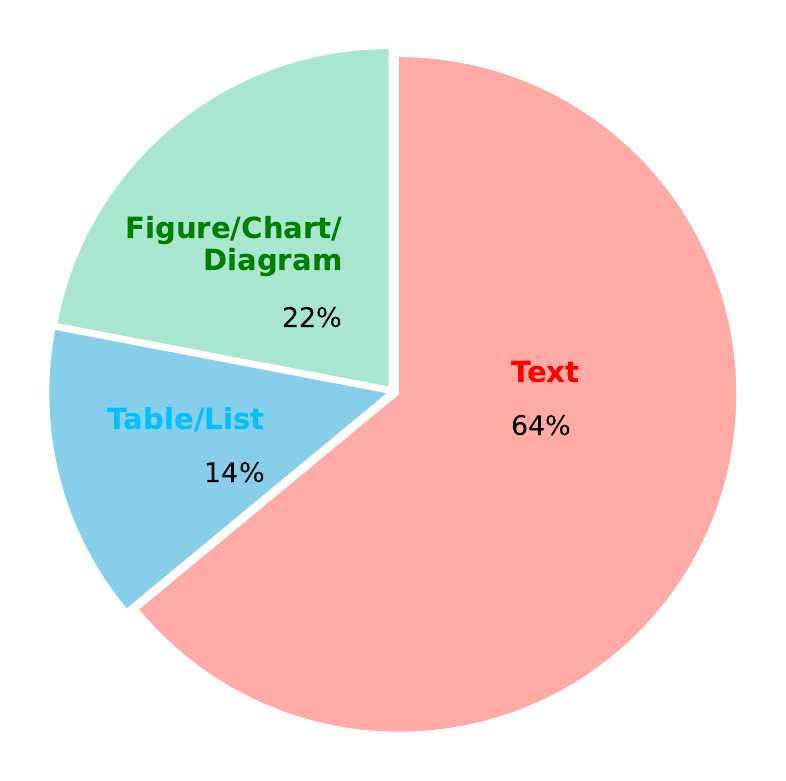}
\caption{Qualitative analysis of the modalities that predominantly contribute to incorrect answers in Doc-REFRAG.}
\label{fig:qualitative_analysis}
\end{figure}

To better understand the failure modes of Doc-REFRAG, we conduct a qualitative analysis on 100 incorrectly answered questions, categorizing the source of errors based on the type of information required for correct inference. As shown in Fig.~\ref{fig:qualitative_analysis}, the majority of errors (64\%) stem from textual content that is dense or semantically complex, where critical details are embedded across multiple paragraphs and require fine-grained reasoning.

In contrast, 22\% of failures involve figures, charts, or diagrams---visual elements that may not be fully reconstructed due to selective chunking, leading to incomplete interpretation. Another 14\% arise from structured data such as tables or lists, which often contain precise numerical or relational information that can be lost during compression if key rows or columns are pruned.

These results suggest that Doc-REFRAG handles unstructured textual content effectively, with the majority of failures (64\%) concentrated in the most challenging regime: dense or semantically complex text requiring fine-grained multi-paragraph reasoning. Visual and structured content together account for the remaining 36\% of errors, reflecting the inherent trade-off between compression efficiency and fidelity when selective chunking omits visually or numerically critical regions.

\section{Empirical Efficiency Validation}
\label{appendix:efficiency}

To quantitatively assess the efficiency gains of Doc-REFRAG, we define the acceleration ratio as:
$$
\text{Acceleration Ratio} = \frac{\text{Latency}_{\text{baseline}}}{\text{Latency}_{\text{ours}}},
$$
where latency is measured in terms of time-to-first-token (TTFT) and time-to-inter-token (TTIT) under a batch size of 1, with $N$ retrieved images conditioned on each question.

\begin{figure}[ht]
\centering
\includegraphics[width=\columnwidth]{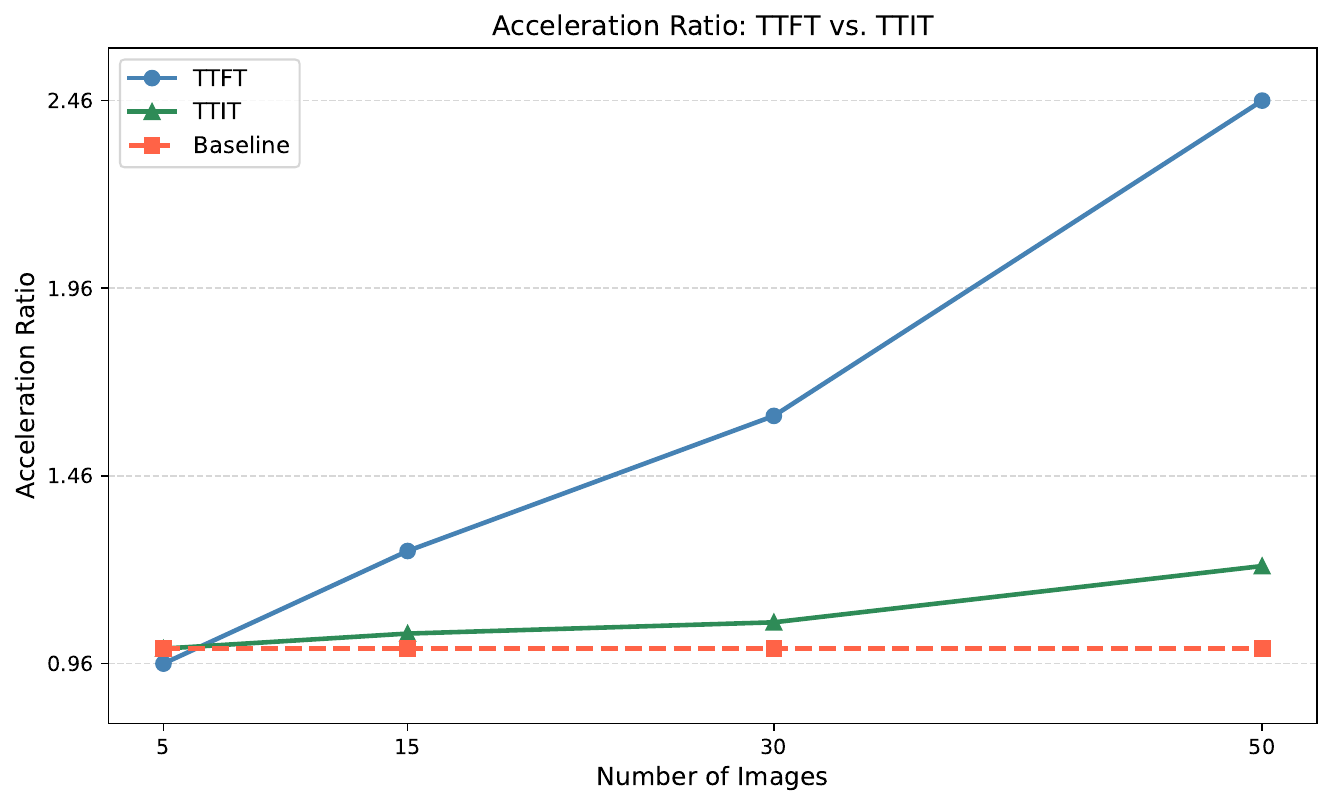}
\caption{Acceleration ratio of Doc-REFRAG versus DocOwl2 in terms of TTFT (blue) and TTIT (green), as a function of the number of retrieved images. The red dashed line indicates the baseline (DocOwl2 Latency = 1$\times$).}
\label{fig:acceleration_ratio}
\end{figure}

As shown in Fig.~\ref{fig:acceleration_ratio}, the TTFT acceleration ratio increases monotonically with $N$, peaking at 2.46$\times$ when $N=50$. This reflects the effectiveness of our question-guided chunk selection in reducing visual tokens, thereby accelerating initial attention computation. In contrast, TTIT acceleration grows gradually (max 1.12$\times$), indicating that post-initial-token processing still involves selected chunks. Moreover, Doc-REFRAG achieves an average accuracy of 50.2\% with 50 retrieved images, substantially outperforming DocOwl2, which attains only 37.2\%.

\section{Disentangling Architectural and Data Contributions}
\label{appendix:disentangle}

We next analyze the independent and joint contributions of architectural design and training data to Doc-REFRAG's overall performance. To this end, we construct a $2 \times 2$ factorial analysis using results already reported in Tab.~\ref{tab:dataset_ablation}, varying both architecture (DocOwl2 vs.\ Doc-REFRAG) and training data (without vs.\ with DocLongRAG).

\begin{table}[ht]
\centering
\resizebox{\columnwidth}{!}{%
\begin{tabular}{lcc}
\toprule
 & \textbf{w/o DocLongRAG} & \textbf{w/ DocLongRAG} \\
\midrule
DocOwl2 (base architecture) & 45.8 & 49.7 \\
\rowcolor{gray!10}
\textbf{Doc-REFRAG (our architecture)} & 48.9 & \textbf{55.1} \\
\midrule
Data gain ($\Delta_\text{data}$) & \multicolumn{2}{c}{$49.7 - 45.8 = +3.9\%$} \\
Architecture gain ($\Delta_\text{arch}$) & \multicolumn{2}{c}{$48.9 - 45.8 = +3.1\%$} \\
Synergy ($\Delta_\text{syn}$) & \multicolumn{2}{c}{$55.1 - 49.7 - 3.1 = +2.3\%$} \\
Total gain & \multicolumn{2}{c}{$55.1 - 45.8 = +9.3\%$} \\
\bottomrule
\end{tabular}%
}
\vspace{4pt}
\caption{$2{\times}2$ factorial decomposition of average accuracy (\%) across DUDE, Vidore, and MMDocIR. Base: DocOwl2 (orig.). $\Delta_\text{data}$: gain from DocLongRAG alone. $\Delta_\text{arch}$: gain from Doc-REFRAG architecture alone. $\Delta_\text{syn}$: additional gain from their interaction.}
\label{tab:disentangle}
\end{table}

Tab.~\ref{tab:disentangle} reveals three complementary sources of improvement.

\paragraph{Data contribution ($+3.9\%$).} Fine-tuning the original DocOwl2 on DocLongRAG alone raises average accuracy from 45.8\% to 49.7\%. This is consistent with the finding in Sec.~\ref{sec:dataset_ablation} that DocLongRAG provides a broadly beneficial training signal for models across different architectures.

\paragraph{Architecture contribution ($+3.1\%$).} Even without DocLongRAG, replacing DocOwl2's encoding strategy with Doc-REFRAG's chunk compression and RL-based selector raises accuracy from 45.8\% to 48.9\%. This demonstrates that the architectural design itself provides meaningful gains that are independent of the new training data.

\paragraph{Synergistic interaction ($+2.3\%$).} The full system (Doc-REFRAG trained on DocLongRAG) achieves 55.1\%, exceeding the sum of the two individual effects by 2.3 percentage points. This super-additive interaction arises because Doc-REFRAG's question-guided compression makes the model better able to exploit the noisy-retrieval training signal present in DocLongRAG: the RL selector learns to identify relevant chunks from heterogeneous retrieved images, a skill that DocLongRAG specifically exercises.

In summary, both contributions are independently meaningful and mutually reinforcing. The architectural gains cannot be explained by data alone: even when DocOwl2 is fine-tuned on DocLongRAG (49.7\%), it remains 5.4 percentage points below Doc-REFRAG Full (55.1\%), demonstrating the independent value of our architectural design. Conversely, the data gains cannot substitute for the architecture: Doc-REFRAG without DocLongRAG (48.9\%) still lags behind the full system (55.1\%), confirming that the dataset and the architecture are complementary rather than redundant.

\section{Validating the DocLongRAG Construction Pipeline}
\label{appendix:naive_baseline}

We investigate whether the performance improvements attributed to DocLongRAG stem from the carefully engineered construction pipeline (hard negative mining and quality filtering) or simply from exposing the model to a larger number of images during training. To isolate these factors, we design a naive random baseline that uses the identical model architecture but replaces the curated DocLongRAG training data with randomly assembled training instances.

Specifically, for this baseline---denoted \textbf{Doc-REFRAG (naive random)}---we draw questions at random from Doc-750K and OpenDocVQA, and pair each question with approximately 37 randomly sampled images (matching the average input length in DocLongRAG), without performing any hard negative mining or quality filtering. All other training hyperparameters and architectural choices are held constant.

\begin{table}[ht]
    \centering
    \resizebox{\columnwidth}{!}{
    \begin{tabular}{l|ccc|c}
        \toprule
        \textbf{Model} & \textbf{DUDE} & \textbf{Vidore} & \textbf{MMDocIR} & \textbf{Average} \\
        \midrule
        \underline{DocOwl2 (orig.)}            & \underline{44.7} & \underline{47.5} & \underline{45.2} & \underline{45.8} \\
        Doc-REFRAG (naive random)  & 34.2 & 36.1 & 35.8 & 35.4 \\
        \rowcolor{gray!10}
        \textbf{Doc-REFRAG (Full)} & \textbf{53.2} & \textbf{57.3} & \textbf{54.9} & \textbf{55.1} \\
        \bottomrule
    \end{tabular}
    }
    \vspace{4pt}
    \caption{Comparison of Doc-REFRAG trained on randomly assembled data (naive random) versus the full DocLongRAG pipeline. The naive random baseline uses the same architecture but pairs each question with $\sim$37 randomly selected images, without hard negative mining or quality filtering. Results are reported as average accuracy (\%~$\uparrow$) on DUDE, Vidore, and MMDocIR. See also Tab.~\ref{tab:dataset_ablation}.}
    \label{tab:naive_baseline}
\end{table}

Tab.~\ref{tab:naive_baseline} yields three clear findings. First, Doc-REFRAG (naive random) achieves only 35.4\% average accuracy---10.4 percentage points \emph{below} DocOwl2 (orig.)\ at 45.8\%---demonstrating that simply exposing the model to more images is not only unhelpful but actively harmful: random image pairings introduce retrieval noise that disrupts the model's reasoning without providing a useful training signal. Second, the full DocLongRAG pipeline raises average accuracy from the DocOwl2 baseline of 45.8\% to 55.1\%, a gain of 9.3 percentage points, confirming that the hard negative mining and quality filtering steps are the true source of improvement. Third, the gap between Doc-REFRAG (Full) and Doc-REFRAG (naive random) is 19.7 percentage points (55.1\% $-$ 35.4\%), clearly isolating the contribution of the \emph{pipeline design} from the contribution of raw image quantity. Together, these results validate that the DocLongRAG construction pipeline itself, not mere data volume, is responsible for the observed performance gains.

\end{document}